\documentclass[10pt,conference]{IEEEtran}
\IEEEoverridecommandlockouts
\usepackage{cite}
\usepackage{amsmath,amssymb,amsfonts}
\usepackage{algorithmic}
\usepackage{graphicx}
\usepackage{textcomp}
\usepackage{xcolor}
\def\BibTeX{{\rm B\kern-.05em{\sc i\kern-.025em b}\kern-.08em
    T\kern-.1667em\lower.7ex\hbox{E}\kern-.125emX}}

\definecolor{NewBlue}{rgb}{0.1, 0.1, 0.7}
\definecolor{NewRed}{rgb}{0.7, 0.1, 0.1}
\usepackage[colorlinks=true,
    linkcolor=NewBlue,
    citecolor=magenta,
    urlcolor=cyan]{hyperref}

\newcommand{\memento}[1]{}

\begin{document}

\title{Quantum Circuit Switching\\with One-Way Repeaters in Star Networks}

\author{\IEEEauthorblockN{Álvaro G. Iñesta\IEEEauthorrefmark{1}\IEEEauthorrefmark{2}\IEEEauthorrefmark{3},
Hyeongrak Choi\IEEEauthorrefmark{4},
Dirk Englund\IEEEauthorrefmark{4}, and
Stephanie Wehner\IEEEauthorrefmark{1}\IEEEauthorrefmark{2}\IEEEauthorrefmark{3}}
\IEEEauthorblockA{\IEEEauthorrefmark{1}QuTech, Delft University of Technology, Lorentzweg 1, 2628 CJ Delft, The Netherlands}
\IEEEauthorblockA{\IEEEauthorrefmark{2}Quantum Computer Science, EEMCS, Delft University of Technology, Mekelweg 4, 2628 CD Delft, The Netherlands}
\IEEEauthorblockA{\IEEEauthorrefmark{3}Kavli Institute of Nanoscience, Delft University of Technology, Lorentzweg 1, 2628 CJ Delft, The Netherlands}
\IEEEauthorblockA{\IEEEauthorrefmark{4}Research Laboratory of Electronics, Massachusetts Institute of Technology, Cambridge, Massachusetts 02139, USA}
\thanks{Corresponding author: Álvaro G. Iñesta (a.gomezinesta@tudelft.nl).}}

\maketitle

\begin{abstract}
Distributing quantum states reliably among distant locations is a key challenge in the field of quantum networks. One-way quantum networks address this by using one-way communication and quantum error correction. 
Here, we analyze quantum circuit switching as a protocol to distribute quantum states in one-way quantum networks.
In quantum circuit switching, pairs of users can request the delivery of multiple quantum states from one user to the other.
After waiting for approval from the network, the states can be distributed either sequentially, forwarding one at a time along a path of quantum repeaters, or in parallel, sending batches of quantum states from repeater to repeater.
Since repeaters can only forward a finite number of quantum states at a time, a pivotal question arises: is it advantageous to send them sequentially (allowing for multiple requests simultaneously) or in parallel (reducing processing time but handling only one request at a time)?
We compare both approaches in a quantum network with a star topology.
Using tools from queuing theory, we show that requests are met at a higher rate when packets are distributed in parallel, although sequential distribution can generally provide service to a larger number of users simultaneously.
We also show that using a large number of quantum repeaters to combat channel losses limits the maximum distance between users, as each repeater introduces additional processing delays.
These findings provide insight into the design of protocols for distributing quantum states in one-way quantum networks.
\end{abstract}

\begin{IEEEkeywords}
circuit switching, resource allocation, one-way, quantum network, quantum repeaters
\end{IEEEkeywords}

\section{Introduction}
In the field of quantum networking, the efficient delivery of quantum information between distant users remains a central challenge \cite{ruf2021quantum}.
A common strategy is to employ quantum repeaters to facilitate quantum communication among remote network nodes \cite{Munro2015,Wehner2018}.
Many repeater architectures require two-way communication and long-lived quantum memories, as they rely on the distribution of entanglement via heralded generation \cite{Barrett2005,Bernien2013} and entanglement swaps \cite{Zukowski1993,Duan2001,Sangouard2011}.
Conversely, the so-called \emph{third-generation quantum repeaters} use quantum error correction to send quantum data using only one-way communication \cite{munro2012quantum, Munro2015, muralidharan2014ultrafast, borregaard2020one}.
Here, we focus on one-way quantum networks, which employ third-generation repeaters to distribute quantum states.

\begin{figure*}[t!]
    \centering
    \includegraphics[width=0.9\linewidth]{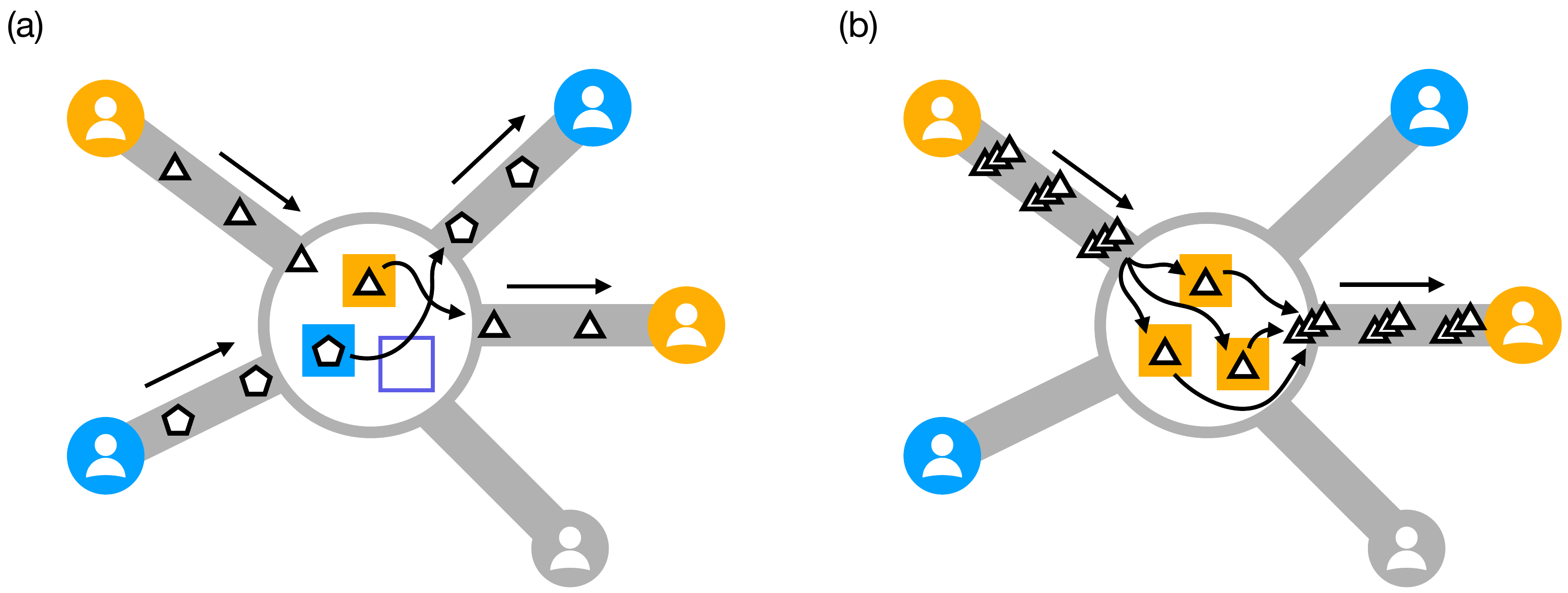}
    \caption{\textbf{In QCS, packets can be distributed  (a) sequentially or (b) in parallel.}
    Illustration of a QCS scheme with (a) sequential and (b) parallel distribution of packets, in a star network with five users and a single central repeater (big circle).
    The repeater contains $k=3$ forwarding stations (squares), therefore it can forward at most $k=3$ quantum data packets (triangles and pentagons) at a time.
    In sequential distribution of packets, a single forwarding station is reserved to meet each of the requests submitted by the pairs of users in orange and blue. In parallel distribution, all three stations are used in parallel to meet a single request at a time.
    }
    \label{fig.qcs}
\end{figure*}

In one-way quantum networks, each quantum state that must be delivered to a remote location (e.g., a data qubit or half of an entangled pair) is encoded and sent over the network as a \emph{quantum data packet}, which consists of the encoded state and some additional metadata (such as the packet destination) \cite{diadamo2022packet}.
In classical networks, data packets are commonly distributed across the network according to a packet switching or a circuit switching protocol \cite{Liew2010}.
In packet switching \cite{Baran1964,Davies1979}, users can send data packets at will, which are then forwarded from one router to the next one whenever possible.
Routers may store the packets until they are able to forward them.
In circuit switching (see, e.g., \cite{Broomell1983}), users first request service to the network. Later, the network reserves one or multiple paths (also called circuits) where routers reserve dedicated resources to meet that request, and there is in principle no need to store data in intermediate routers.
In classical networks, packet switching has found more success than circuit switching since it makes better use of the resources in large networks with conflicting routes and high traffic loads.
In quantum networks, quantum data packets can also be distributed according to a packet switching \cite{diadamo2022packet} or a circuit switching protocol.
Quantum circuit switching has been studied in the context of two-way quantum repeaters \cite{Aparicio2011}, but not in the context of one-way quantum networks, to the best of our knowledge.

The main challenge when designing a one-way protocol to distribute quantum states is decoherence: quantum states have a limited lifetime and cannot be stored indefinitely in intermediate quantum repeaters.
This means that the protocol must ensure rapid delivery of the quantum data packets.
Additionally, some multi-party applications may need to consume multiple quantum states simultaneously (e.g.,
to implement verifiable blind quantum computations, multiple qubits must be sent from a client to a server, where they must coexist while some operations are performed \cite{Leichtle2021, Davies2023}).
In those cases, the quantum states must be distributed within some time window -- otherwise, by the time the last state is distributed, the quality of the first state would have decayed too much due to decoherence.
Consequently, one-way protocols must prevent large variations in the time between the delivery of successive quantum data packets to ensure all of them are delivered within a specific time window.
    We expect the pre-allocation of resources in circuit switching to enable the protocol to meet these timing constraints required to successfully distribute quantum states over a network, regardless of network traffic.
In this work, we introduce \emph{quantum circuit switching} (QCS) for one-way quantum networks, and we investigate how to allocate resources for each request to distribute quantum states.



    Each quantum repeater has limited capabilities, namely, it can only forward a finite number of packets, $k$, at a time (see Figure~\ref{fig.qcs}).
    We then say that the repeater has $k$ \emph{forwarding stations}, which operate independently and can forward one packet at a time.
    When multiple requests need to use the same repeater, the circuit switching protocol can either ($i$) reserve one forwarding station per request and process up to $k$ requests simultaneously (Figure~\ref{fig.qcs}a), or ($ii$) reserve all $k$ forwarding stations to meet one request at a time (Figure~\ref{fig.qcs}b). For simplicity, we omit the intermediate case where more than one but less than $k$ stations are reserved for each request.
    In the first approach, quantum data packets are distributed \emph{sequentially}, while in the second approach packets are distributed in \emph{parallel} (requiring some form of frequency multiplexing).
    
    The performance of both approaches can be measured using the \emph{mean sojourn time}, which is the time passed since the request is submitted until it is met -- this includes some \emph{waiting time} before the request is processed and the \emph{service time} needed to successfully distribute all the quantum packets.
    Packets can be lost while traveling from node to node, so the total number of packets that need to be sent to successfully deliver a given number of packets is a priori unknown. As a consequence, the sojourn, waiting, and service times are random variables.
    We expect shorter service times when packets are distributed in parallel rather than sequentially. However, parallel distribution can only serve one request at a time, and this might entail large waiting times.
    A natural question arises: \emph{in what situations is it actually advantageous to distribute quantum packets in parallel instead of sequentially?} That is, \emph{when does parallel distribution provide shorter mean sojourn times?}
    Addressing this resource-allocation question early on is crucial for the design of efficient protocols for distributing quantum states within one-way quantum networks.
    Here, we answer the question assuming a star network: a simplified topology where all users are connected via a number of quantum repeaters to a central repeater (similar to a quantum switch architecture \cite{Vardoyan2021,Vardoyan2021b}).

Our main contributions are the following:
\begin{itemize}
    \item We formalize the concept of quantum circuit switching in the context of one-way quantum networks.
    \item We provide analytical tools to compute the mean sojourn time of a QCS scheme, using techniques from queuing theory.
    \item We compare QCS schemes with sequential and parallel distribution of packets in a star network, and provide heuristics for the design of QCS protocols in more complex networks.
\end{itemize}

Our main findings are the following:
\begin{itemize}
    \item In star networks with sequential distribution of packets, we can increase the number of forwarding stations $k$ to increase the number of users supported by the network (i.e., the maximum number of users that allow requests to be met within a finite amount of time), which scales as $\sim \sqrt{k}$. This is not possible in parallel distribution: the number of users is capped and cannot be increased indefinitely by adding more forwarding stations per repeater.
    \item Parallel distribution generally provides smaller mean sojourn times than sequential distribution.
    \item There exists a trade-off between the number of users and the physical distances between them:
    too many users and too long distances overload the system and yield infinite waiting times.
    \item When there is a large number of users, adding more intermediate repeaters to combat channel losses and minimize the probability of packet loss does not allow the users to be located further away. In fact, adding more repeaters limits the size of the network to smaller scales, since each repeater introduces additional delays.
\end{itemize}

In Section \ref{sec.setup}, we explain the problem setup: the quantum network and quantum data packets model, the requests model, and the quantum circuit switching scheme.
In \ref{sec.results}, we present our main results: we analyze the operating regimes of both packet distribution strategies and compare their performance. We do this in a variety of scenarios, considering different quantum repeater architectures based on the financial budget available.
Lastly, in \ref{sec.outlook}, we discuss the limitations and the implications of our work.

\vspace{10pt}
\section{Problem setup}\label{sec.setup}
In this Section, we present the problem setup.
In \ref{sec.model_network}, we discuss our model of quantum network and quantum data packets.
In \ref{sec.model_requests}, we explain how the nodes submit requests and how to measure the performance of the network in meeting those requests.
Lastly, we formalize the concept of quantum circuit switching in \ref{sec.model_qcs}.
We provide a summary of all the parameters introduced in this Section in Table \ref{tab.variables}.

\renewcommand{\arraystretch}{1.2}
\begin{table}[t]
    \centering
    \caption{Parameters of a quantum network running a QCS protocol.}\label{tab.variables}
    \vspace{-2mm} 
\begin{tabular}{cp{0.8\columnwidth}}
\multicolumn{2}{c}{\textbf{Physical topology (star network)}}\\
\hline
	$u$ & Number of users \\
	$L$ & Distance between each user and the central repeater \\
     $N$ & Number of repeaters between each user and the central repeater\\[5pt]
\multicolumn{2}{c}{\textbf{Hardware}}\\
\hline
	$k$ & Number of forwarding stations per repeater\\
	$t_\mathrm{fwd}$ & Forwarding time per forwarding station and quantum data packet\\
 	$p$ & Probability of successful packet delivery from user to user\\
	$c$ & Speed of light in the physical channels\\[5pt]
\multicolumn{2}{c}{\textbf{Requests}}\\
\hline
	$n$ & Number of quantum data packets per request \\
	$w$ & Request time window \\
 	$\lambda_0$ & Request submission rate per pair of users \\[5pt]
\multicolumn{2}{c}{\textbf{Quantum Circuit Switching}}\\
\hline
	$m$ & Number of packets of the same request distributed concurrently ($m=1$ for sequential distribution and $m=k$ for parallel distribution) \\[5pt]
\end{tabular}

\end{table}

\subsection{Quantum networks and quantum data packets}\label{sec.model_network}
We consider a quantum network with \emph{third-generation quantum repeaters} \cite{Munro2015,muralidharan2014ultrafast,Muralidharan2016,borregaard2020one}.
These type of repeaters use quantum error-correction to distribute quantum information using one-way communication, as opposed to first- and second-generation repeaters, which use heralded entanglement generation and two-way communication \cite{Muralidharan2016}.
To forward quantum data, a repeater must first decode an incoming logical state (generally shared as a collection of photons) and correct any errors, then reencode the state again into multiple physical qubits, and finally send the encoded state to the next repeater.

\begin{figure}[t!]
    \centering
    \includegraphics[width=0.9\linewidth]{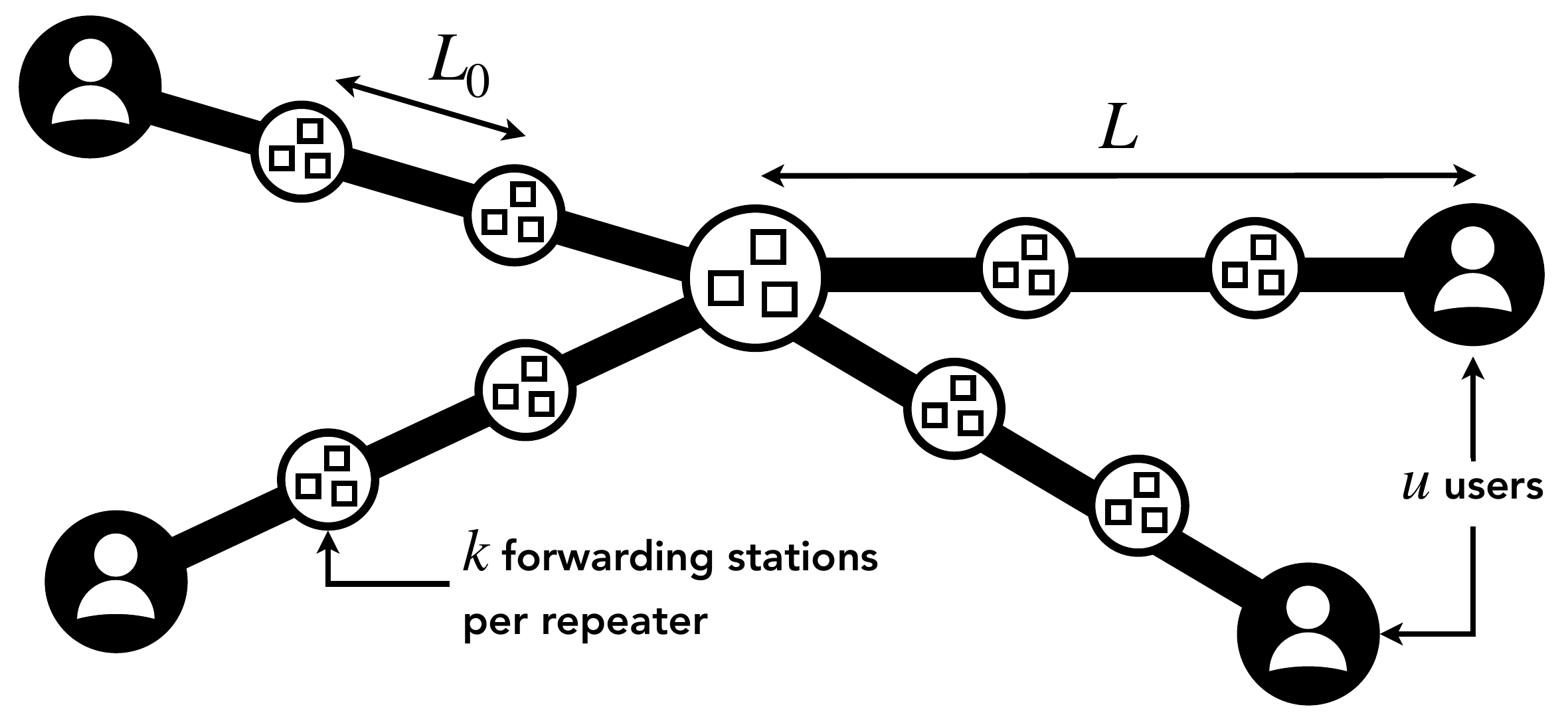}
    \caption{\textbf{Illustration of a star quantum network.}
    $u$ users are at distance $L$ from a central repeater. There are $N$ repeaters between each user and the central repeater ($N=2$ in the figure), and each of them has $k$ forwarding stations (squares). The spacing between adjacent nodes is $L_0 = L/(N+1)$.
    }
    \label{fig.starnetwork}
\end{figure}

We make the following assumptions about the quantum network:
\begin{itemize}
    \item We consider a quantum network with a \emph{star topology}, where $u$ users are connected to a central repeater (see Figure~\ref{fig.starnetwork}). This is the simplest case of network where paths between users intersect, leading to shared resources and routing conflicts.

    \item For simplicity, we assume that all users are at the same physical distance $L$ from the central repeater (see Figure~\ref{fig.starnetwork}). Moreover, there are $N$ repeaters between each user and the central repeater. All repeaters are placed at distance $L_0 = L/(N+1)$ from each other.

    \item Users can encode and decode quantum data packets, according to some quantum code.
    A \emph{quantum data packet} consists of a logical qubit and a classical header containing some metadata, as proposed in ref. \cite{diadamo2022packet}.
    The logical qubit consists of multiple physical qubits, generally in the form of photons.
    The classical header provides relevant information about the routing of the data packet (e.g., the destination of the packet).

    \item Each repeater has $k$ \emph{forwarding stations} that can decode and encode according to the same quantum code.
    Each station can receive/send physical qubits from/to any node.
    Moreover, each station can only forward (i.e., decode, reencode, and send) one quantum data packet at a time, which takes time $t_\mathrm{fwd}$.
    After forwarding a packet, the station can be immediately used to forward another packet from any source node to any destination node (i.e., there is no downtime).

    \item Physical qubits can suffer from noise in the encoding and decoding circuits and in the physical channels. We assume that each packet can be successfully decoded at destination with probability $p$ (this includes the case in which the encoded qubits suffered no errors and also the case in which they suffered a correctable number of errors). With probability $1-p$, a number of uncorrectable errors will be detected at destination and the packet will be discarded.
    We consider the probability of a logical error being unnoticed to be negligible, since we assume the main source of errors to be photon loss in the channels, which can be detected.
\end{itemize}

\subsection{Requests and performance measures}\label{sec.model_requests}
In many quantum network applications, multiple copies of a quantum state need to be distributed over a short interval of time.
For example, verifiable blind quantum computing requires multiple qubits to be sent from a client to a server, where they must coexist while some operations are performed \cite{Leichtle2021,Davies2023}.
Another example is verifiable quantum secret sharing \cite{Crepeau2002}, where the parties involved want to verify that the dealer successfully distributed a quantum secret among them. This verification requires the dealer to distribute multiple ancillary quantum states to each party that will be consumed simultaneously (note that other proposals perform the verification stage sequentially \cite{Lipinska2020}).
In these examples, a number of quantum states must be sent from one network node to another over a short interval of time due to decoherence -- otherwise, when the last state is distributed, the quality of the state that was distributed first would have decayed too much and would not be useful anymore.

With this type of general application in mind, we define an \emph{$(n,w)$-request} as request for $n$ quantum data packets to be distributed among two users within time $w$. Each packet contains a copy of the same quantum state.

We make the following additional assumptions about the requests:
\begin{itemize}
    \item To simplify the analysis, we assume that all pairs of users make requests with the same parameters $n$ and $w$.

    \item Each pair of nodes submits requests following a Poisson process with rate $\lambda_0$. The total \emph{request submission rate} is then
    \begin{equation}
        \lambda = {u \choose 2} \lambda_0 = \frac{u(u-1)}{2}\lambda_0,
    \end{equation}
    where $u$ is the number of users.
    
\end{itemize}

The performance of protocols for quantum state distribution is typically measured with the quality of the distributed states and/or the distribution rate (see, e.g., \cite{Pant2019,Khatri2022,Avis2022,Inesta2023}).
In our model, quantum data packets are either successfully distributed or a failure flag is raised, and therefore there is no need to study the quality of the state.
The main quantity of interest in our problem is therefore the rate at which $(n,w)$-requests are met.
In particular, to measure the performance of the protocol, we propose the \emph{mean sojourn time}: the average time passed since a request is submitted until it is met (i.e., until the $n$-th quantum packet is successfully delivered within the time window $w$).
The sojourn time of a request can be computed as the sum of two main contributions:
\begin{itemize}
    \item \emph{Waiting time} ($T_\mathrm{wait}$): time since the request is submitted until some forwarding stations are available for the request. The waiting time depends on the number of requests in the queue at a given time and is a random variable.

    \item \emph{Service time} ($T_\mathrm{service}$): time since the stations are available for the request until the request is met. Since packets can be lost, the service time is a random variable.
\end{itemize}

In particular, we are interested in the mean sojourn time, which can be computed as
\begin{equation}\label{eq.MST_wait_service}
    \mathbb{E}\left[ T_\mathrm{sojourn} \right] = \mathbb{E}\left[T_\mathrm{wait}\right] + \mathbb{E}\left[T_\mathrm{service}\right].
\end{equation}
In this work, we implicitly refer to the steady-state mean values when we discuss the mean sojourn time, the mean waiting time, and the mean service time. The steady state corresponds to the status of the system where mean values have reached an equilibrium following a transient period.
In the calculation of the sojourn time, one could also add some control time, which would include the time required to submit a request.
We can assume this control time is independent of the choice of protocol for quantum state distribution, and therefore we leave it outside of our analysis (if we assume the control time is constant, our results would just shift towards larger sojourn times).
    
We now summarize how to compute each term from~(\ref{eq.MST_wait_service}):
\begin{itemize}
    \item To compute the mean waiting time, we model the system as an M/G/$s$ queue -- this is a queue model where request arrivals follow a Poisson distribution and therefore are Markovian (M), the service time follows a general distribution (G), and a maximum of $s$ requests can be processed simultaneously.
    This formulation allows us to derive analytical solutions and approximations for the mean waiting time in a variety of situations, although in some cases Monte Carlo methods are required. See Appendices~\ref{app.analytical} and \ref{app.coeff_of_var} for further details.
    
    \item The service time distribution is hard to compute analytically. The reason is that calculating the number of packets that need to be sent until $n$ are successfully received over a time window $w$ is a challenging task \cite{Davies2023}. When packets are sent in batches of $m$ (with $m\leq k$, since repeaters can only forward at most $k$ packets simultaneously) and each packet can be lost with probability $p$, the number of batches until success is denoted by the random variable $B_{n,w,p,m}$.
    The mean service time can be computed in terms of the mean value of $B_{n,w,p,m}$:
    \begin{equation}\label{eq.mean_service}
    \mathbb{E}\left[T_\mathrm{service}\right] = \frac{2L}{c} + t_\mathrm{fwd}\left(2N + \mathbb{E}\left[B_{n,w,p,m}\right]\right),
    \end{equation}
    where $c$ is the speed of light in the communication channels.
    The first term corresponds to the time required for a packet (or a batch of packets) to travel from one user to another.
    The second term accounts for the total delay introduced by the forwarding stations, which depends on the number of repeaters, $N$, and the expected number of batches required, $\mathbb{E}\left[B_{n,w,p,m}\right]$.
    In Appendix~\ref{subsec.servicetime}, we provide a derivation of (\ref{eq.mean_service}), and we also explain how to compute the probability distribution of $B_{n,w,p,m}$ (in some cases, numerical approximations are required).
\end{itemize}

Another relevant quantity is the \emph{load} of the system, which is the ratio between the request submission rate, $\lambda_0 u(u-1)/2$, and the total rate at which requests are serviced, $k/(m\mathbb{E}\left[T_\mathrm{service}\right])$, where $1/\mathbb{E}\left[T_\mathrm{service}\right]$ is the rate at which one request is serviced and $k/m$ is the number of requests that can be processed simultaneously. The load can be written as
\begin{equation}\label{eq.load}
    \rho = \frac{\lambda_0 u (u-1) m}{2k} \mathbb{E}\left[T_\mathrm{service}\right].
\end{equation}
It can be shown that, if and only if $\rho > 1$, the system is overloaded and the sojourn times go to infinity (see Chapter 14 from \cite{VanMieghem2014}).
This happens when requests are submitted at a higher rate than they are serviced.
For example, if the number of users is too large, the number of requests waiting to be serviced will grow indefinitely and the sojourn times will go to infinity over time.
Therefore, service will only be possible if the load remains below one.

\subsection{Quantum Circuit Switching}\label{sec.model_qcs}
Quantum circuit switching (QCS) protocols are a particular type of protocol for on-demand distribution of quantum states.
In QCS, each pair of users can submit $(n,w)$-requests to the network controller.
Incoming requests are placed in a first-in-first-out queue, where older requests have priority over more recent ones.
Once enough resources are available, a request can leave the queue and some network resources are reserved to meet that request.
In our model, the resources reserved are forwarding stations at quantum repeaters over a path that connects both users.
After this, one of the users can start sending quantum data packets, which are forwarded from repeater to repeater until arriving at the other user.
This process continues until $n$ packets are successfully delivered over a sliding time window $w$ (recall that each packet has a probability $p$ of being flagged with an uncorrectable error).
When the process is completed, the forwarding stations become available and can be assigned to the first request in the queue.
Note that we consider a queue with infinite capacity, as opposed to classical circuit switching, where the queue has a maximum size and additional requests are rejected (e.g., new calls were rejected when all lines were occupied in early analog telephone networks) -- we do this because we are interested in understanding the behavior of the system without such a constraint.

The main degree of freedom in our QCS proposal is the number of forwarding stations that are reserved for each request:
\begin{enumerate}
    \item If we reserve only one forwarding station per request, packets must be sent sequentially, i.e., $m=1$, since each station can only forward one packet at a time (Figure \ref{fig.qcs}a). We call this strategy \emph{sequential distribution} of packets.
    \item An alternative is to reserve multiple forwarding stations, such that multiple packets can be routed in parallel. We call this approach \emph{parallel distribution} of packets. For simplicity, we assume that QCS with parallel distribution reserves all $k$ stations for each request (Figure \ref{fig.qcs}b), i.e., $m=k$.
    In practice, this can be done with some form of frequency multiplexing \cite{Collins2007}.
\end{enumerate}
In sequential distribution, the service time $T_\mathrm{service}$ is larger, but many requests can be processed simultaneously (as long as there are $k>1$ forwarding stations), which might yield a shorter waiting time $T_\mathrm{wait}$.
Conversely, parallel distribution provides a faster service (i.e., smaller $T_\mathrm{service}$) at the expense of processing one request at a time, which could entail a longer waiting time $T_\mathrm{wait}$.
It is therefore nontrivial to determine a priori whether packets should be distributed sequentially or in parallel.
The first step in the design of efficient QCS protocols is to answer the following question: \emph{should quantum data packets be distributed sequentially or in parallel?}
This is the main question we address in this work.

\vspace{10pt}
\section{Sequential vs parallel distribution of quantum data packets}\label{sec.results}
In this Section, we explore the situations in which sequential distribution of packets may be advantageous over parallel distribution and vice versa.
We consider two use cases:
\begin{enumerate}
    \item \emph{Small budget}.
    We consider the more affordable one-way quantum repeater architecture from ref.~\cite{borregaard2020one}, which only requires two matter qubits and a single-photon emitter per forwarding station.
    The encoding used by this repeater is not fault-tolerant and therefore can only be used over short inter-repeater distances ($L_0\sim1$ km).
    Over such a short distance, the scheme is near-deterministic, i.e., we can assume packets are successfully delivered with probability $p\approx 1$.
    Since we consider a tight financial budget, we also assume that there is a single central repeater that is directly connected to the users (i.e., $N=0$), which means that the size of the network cannot be larger than $L \sim 1$~km.

    \item \emph{Large budget.}
    In this case, we employ a more expensive type of forwarding station: the all-photonic proposal from ref. \cite{Niu2022}.
    This architecture can encode quantum states using large distance codes, which allows for distribution of packets over longer inter-repeater distances ($L_0>1$ km).
        Since the financial budget is larger than before, we can place $N$ repeaters in between each user and the central repeater.
        The probability of successful delivery depends on the number of repeaters, $N$, and the size of the network,~$L$:
        \begin{equation}\label{eq.p_allphotonic}
            p(L,N) = 10^{-\frac{\alpha_\mathrm{eff} \left( L_0 \right)}{10} 2L},
        \end{equation}
        where $L_0 = L/(N+1)$ is the inter-repeater distance, and $\alpha_\mathrm{eff}(L_0)$ is an effective attenuation coefficient that depends on the type of encoding used and on the repeater efficiency.
        Here, we assume $\alpha_\mathrm{eff}(L_0) \approx 10^{-6} (277 L_0^2 + 29 L_0^4)$ dB/km, which corresponds to forwarding stations that employ the [[48, 6, 8]] generalized bicycle code \cite{Panteleev2021} and have a 90\% efficiency (which incorporates photon-source and detector efficiencies, on-chip loss, and coupling losses into a single parameter). This encoding and efficiency was also used as an example in ref.~\cite{Niu2022} -- see Appendix \ref{app.allphotonic} for further details.
        To make good use of the budget available, we choose $N$ following the strategy from ref.~\cite{Niu2022}, where the authors propose maximizing the number of repeaters per kilometer divided by the probability $p$, i.e., they optimize the cost function~$(2N+1)/(Lp)$. For the $[[48, 6, 8]]$ code and a fixed distance $L$, the value of $N$ that minimizes the cost is the one that yields $p\approx0.7$ (e.g., the optimal solutions for $L=7.5, 13, 18, 30$~km are $N=0,1,2,5$).
\end{enumerate}

The main motivation behind these two use cases is that they correspond to a scenario with deterministic delivery of packets over short distances ($p=1$, use case 1) and a scenario with probabilistic delivery over longer distances ($p<1$, use case~2).

In the following subsections, we analyze the performance of a QCS protocol in the previous use cases.
In Subsection \ref{subsec.usecases_load} we analyze the maximum number of users that the system can support before sojourn times go to infinity.
Then, in \ref{subsec.usecases_sojourn}, we compare the performance of sequential and parallel distribution, in terms of the mean sojourn time, when the budget is small (use case 1) and when it is large (use case 2).
Lastly, in \ref{subsec.going_far}, we focus on use case 2 and study the interplay between the number of users and the distances between them, considering a fixed number of repeaters $N$.
In Appendix \ref{app.parametervalues}, we motivate the parameter values chosen in the examples from this Section.

\subsection{Critical number of users}\label{subsec.usecases_load}
Before looking at the performance of sequential and parallel distribution of packets, the first question we ask is: \emph{how many users can be supported by the quantum network?} That is, we want to know the maximum number of users that can be serviced before the sojourn times go to infinity -- what we call the \emph{critical number of users}, $u_\mathrm{crit}$. We find $u_\mathrm{crit}$ by ensuring that the load of the system (\ref{eq.load}) is below 1, which yields
\begin{equation}\label{eq.ucrit}
    u_\mathrm{crit} = \left\lfloor \frac{1}{2} + \frac{1}{2} \sqrt{1 + \frac{8k}{\lambda_0 m \mathbb{E}\left[T_\mathrm{service}\right]}} \right\rfloor,
\end{equation}
where $m = 1$ for sequential distribution and $m=k$ for parallel distribution.

\begin{figure}[t!]
    \centering
    \includegraphics[width=0.95\linewidth]{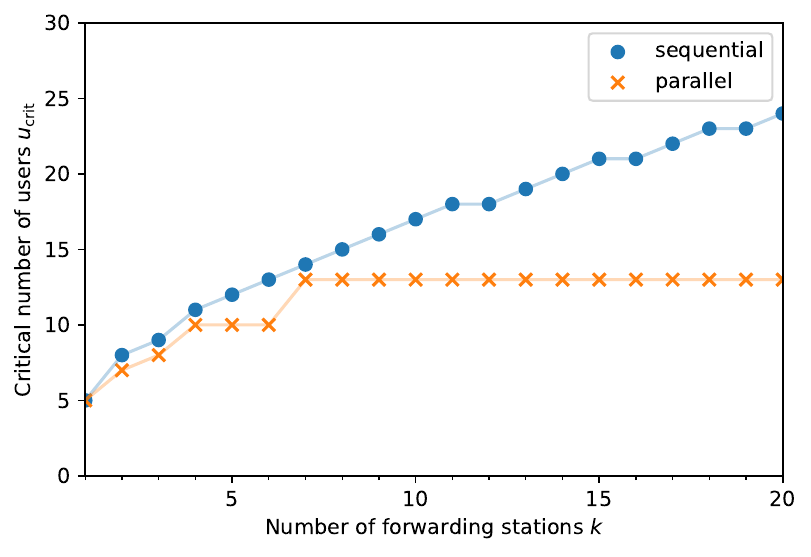}
    \caption{\textbf{A network with more forwarding stations $k$ can support more users, but only when distribution of packets is sequential.}
    Critical number of users, $u_\mathrm{crit}$, vs number of forwarding stations, $k$, for QCS with sequential (blue dots) and parallel (orange crosses) distribution of packets, in the small budget use case (see (\ref{eq.ucrit}) and (\ref{eq.Tservice_p1})).
    The critical number of users is the maximum number of users the system can support before the sojourn time goes to infinity.
    In sequential distribution, $u_\mathrm{crit}(k)$ scales as $\sqrt{k}$. In parallel distribution, $u_\mathrm{crit}(k) = \mathrm{constant}$, $\forall k>n$.
    Parameters used in this figure: $N=0$, $p=1$, $n=7$, $w=10$, $\lambda_0 = 10^{-4}$~$\mu$s$^{-1}$, $c=0.2$ km/$\mu$s, $t_\mathrm{fwd} = 100$~$\mu$s.}
    \label{fig.crit-users-p1}
\end{figure}

In the small-budget situation (use case 1), $p=1$ and $N=0$. Then, the mean service time from (\ref{eq.mean_service}) takes a closed form:
\begin{equation}\label{eq.Tservice_p1}
    \mathbb{E}\left[T_\mathrm{service}\right]_{p=1,N=0} = \frac{2L}{c} + t_\mathrm{fwd} \left\lceil \frac{n}{m} \right\rceil.
\end{equation}
This expression can be derived using (\ref{eq.mean_service}) and the fact that $\mathbb{E}[B_{n,w,1,m}] = \left\lceil n/m \right\rceil$ (for further details on the latter, see Appendix~\ref{subsec.servicetime}).
From (\ref{eq.ucrit}) and (\ref{eq.Tservice_p1}), one can show that, for a small budget, the critical number of users is larger when packets are distributed sequentially rather than in parallel.
This can be seen in Figure \ref{fig.crit-users-p1}, which shows $u_\mathrm{crit}$ for increasing number of forwarding stations.
It is possible to provide service to a larger number of users (i.e., increase $u_\mathrm{crit}$) by increasing the number of forwarding stations $k$, but only when packets are distributed sequentially. In particular, the critical number of users scales as $\sqrt{k}$.
When packets are distributed in parallel, increasing the number of stations beyond $k=n$ does not allow for an increased number of users. This is a consequence of the parallel distribution model: we assume that all the forwarding stations are used simultaneously for a single request. When packet distribution is deterministic ($p=1$), only $n$ forwarding stations are needed to distribute $n$ packets in parallel, and increasing $k$ beyond this number will not provide any added benefit -- those extra forwarding stations will remain unused.

When the financial budget is large (use case 2), the service time follows a nontrivial distribution and the mean value cannot be computed with (\ref{eq.Tservice_p1}) anymore.
Depending on the values of $p$ and $w$, we were able to compute the mean service time analytically or required Monte Carlo sampling (see Appendix~\ref{app.analytical} for further details).
Nevertheless, the scaling of $u_\mathrm{crit}$ with $k$ remains similar to the behavior shown in Fig.~\ref{fig.crit-users-p1} (we provide the same plot for a large budget and different combinations of $N,L,w$ in Appendix~\ref{app.ucrit_probabilistic}).
We observed that, for most parameter regimes, sequential distribution can still provide service to a larger number of users than parallel distribution.
However, this is not true in the parameter regime where resources are scarce, namely, when: ($i$) packet distribution is not deterministic ($p<1$; specifically in our use case, $p \approx 0.7$), ($ii$) applications require many states to be distributed over a short period of time or users have short-lived quantum memories (small $w$ compared to $n$), and ($iii$) there are only a few forwarding stations per repeater (small $k$).
In a situation that meets these three conditions, many states must be successfully delivered over a short time window. Since each packet has a nonzero probability of failure, this process can take a very long time when packets are delivered one by one, as successful packets older than the time window are discarded. Hence, parallel distribution can complete requests much faster and therefore provide service to a larger number of users. This reasoning only holds if $k$ is small: if there is a large number of forwarding stations, more users can be served simultaneously with sequential distribution.

\begin{figure*}[t!]
    \centering
    \includegraphics[width=\linewidth]{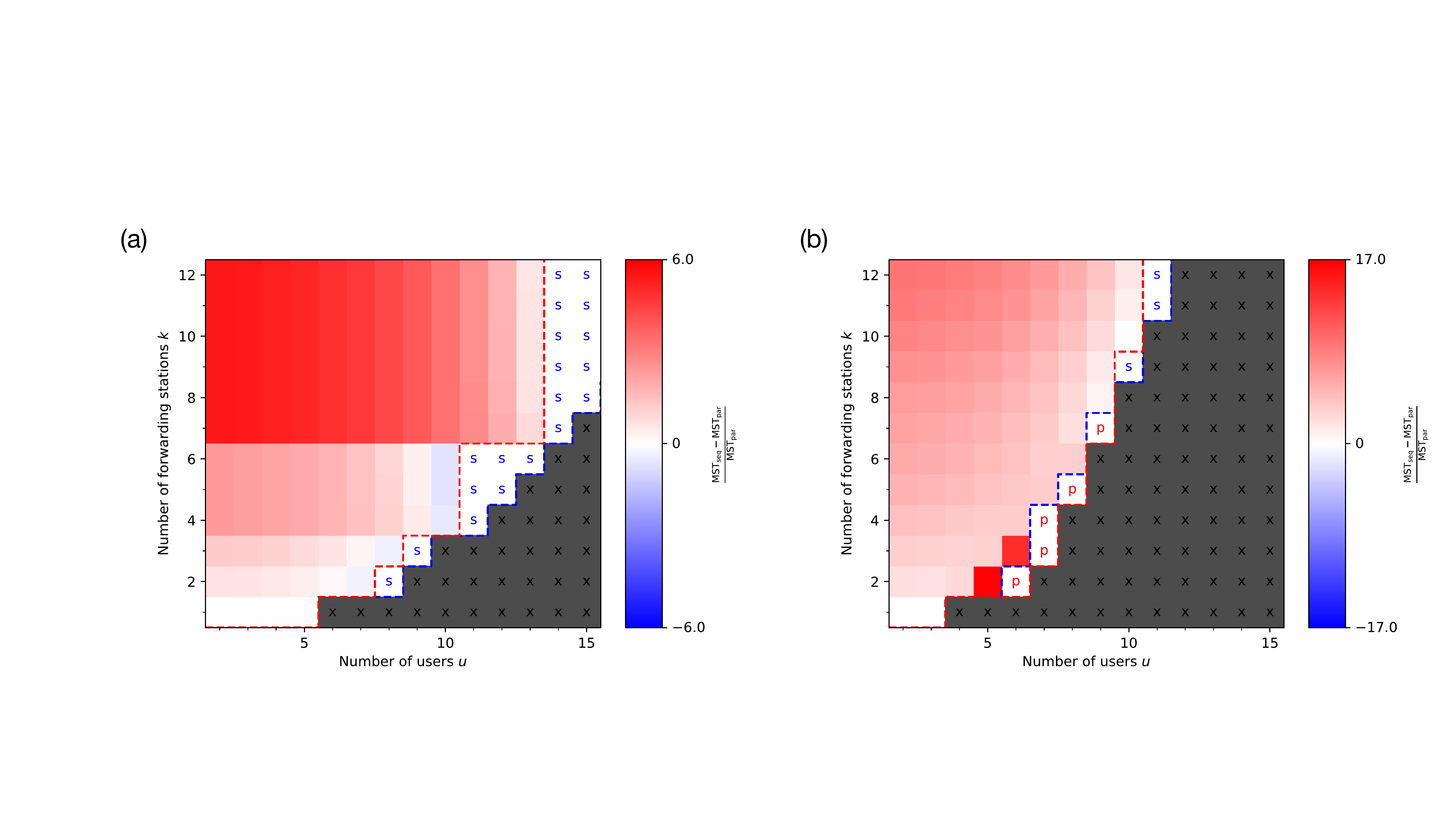}
    \caption{\textbf{Parallel distribution of packets is generally faster.}
    Relative difference in mean sojourn time (MST) between sequential and parallel packet distribution, for different numbers of users $u$ and forwarding stations $k$.
    \textbf{(a)} Small-budget use case ($p=1$; $w\geq n$; $N=0$, $L=1$ km) and \textbf{(b)} large-budget use case ($p\approx0.7$) with $N=0$, $L=7.5$ km, and $w=8$.
    Sequential/parallel distribution provides lower MST in blue/red regions.
    In regions with an `s'/`p', only sequential/parallel distribution can provide service (i.e., yield finite MST).
    In dark regions with an `x', no service is possible.
    Parameters used in this figure: $n=7$, $\lambda_0 = 10^{-4}$~$\mu$s$^{-1}$, $c=0.2$ km/$\mu$s, $t_\mathrm{fwd} = 100$~$\mu$s.
    MSTs in (a) calculated with (\ref{eq.Tservice_p1}). MSTs in (b) calculated with Monte Carlo sampling with $10^4$ samples
    (the standard error in the relative difference in MST was below 0.5 for every combination of parameters).
    }
    \label{fig.MST-usecasesAC}
\end{figure*}

In the large-budget scenario, the asymptotic behavior in $k$ is similar to the small-budget case.
When $p<1$, the mean service time is lower bounded by (\ref{eq.Tservice_p1}), and therefore the critical number of users for $p=1$ (small budget) is an upper bound for $p<1$ (large budget).
In particular, this means that $u_\mathrm{crit}$ cannot scale faster than $\sqrt{k}$ with sequential distribution, and it converges to a constant as $k\rightarrow\infty$ with parallel distribution.
As mentioned earlier, we provide some graphical examples in Appendix~\ref{app.ucrit_probabilistic}.

\subsection{Mean sojourn time}\label{subsec.usecases_sojourn}
Now we focus on the scenarios in which both sequential and parallel distribution can provide service within a finite mean sojourn time (MST).
The MST can be computed according to (\ref{eq.MST_wait_service}), which requires calculating the mean waiting time and the mean service time in advance.
In Appendix~\ref{app.analytical} we provide analytical formulas and numerical methods to compute them.

Figure \ref{fig.MST-usecasesAC} shows the relative difference in MST between sequential and parallel distribution, for the small-budget use case (Figure \ref{fig.MST-usecasesAC}a) and the large-budget use case (Figure \ref{fig.MST-usecasesAC}b).
In both use cases, there is a region where both sequential and parallel distributions are possible. This region is shaded blue/red when sequential/parallel distribution is faster.
When the number of users is too large, service is not possible. We indicate with an `s'/`p' those regions where service can only be provided by sequential/parallel distribution. In dark regions with an `x', service is not possible at all.
We draw the following main observations from the figure:
\begin{itemize}
    \item In the small-budget use case (Figure \ref{fig.MST-usecasesAC}a), increasing the number of users (for fixed $k$) leads to a region where only sequential distribution can provide service. If the number of users keeps increasing, no service is possible at all.
    In the large-budget case (Figure \ref{fig.MST-usecasesAC}b), we observe a similar behavior, except that there is also a region where only parallel distribution (instead of sequential) is possible. This is the same phenomenon that was discussed in the previous section: when $p<1$, $w$ is close to $n$, and $k$ is small, parallel distribution supports more users.
    In Appendix~\ref{app.MST}, we show other examples in which there is a region where only parallel distribution is possible.
    
    \item For a fixed number of forwarding stations $k$, parallel distribution is faster than sequential distribution when the number of users is small.
    As we increase the number of users (for fixed $k$), the advantage of parallel over sequential distribution generally decreases. For some values of $k$, sequential distribution eventually becomes slightly better.
    This happens because the main feature of sequential distribution is that multiple requests can be processed simultaneously, and this is particularly beneficial when there is a large number of requests (in our model, a large number of users implies a large number of requests).
    In some other cases, parallel distribution is always better (e.g., in the small-budget example from Figure~\ref{fig.MST-usecasesAC}a with $k>6$).
    When parallel distribution has a larger critical number of users (e.g., in the large-budget case from Fig \ref{fig.MST-usecasesAC}b with $k=2,3$), the advantage in MST over sequential distribution can actually increase dramatically as the number of users increases (e.g., increasing $u$ from 4 to 5 when $k=2$, in Figure \ref{fig.MST-usecasesAC}b).
    The reason is that the MST of sequential distribution diverges as we get close to a region where only parallel distribution is possible.
    This can be seen more clearly in Appendix~\ref{app.MST}, where we plot the MST vs the number of users for fixed $k$.

    \item The difference in MST converges to a constant value as $k\rightarrow\infty$.
    In fact, the MST should converge to a constant as $k\rightarrow\infty$ for both sequential and parallel distribution.
    Intuitively, once we have enough forwarding stations to meet all incoming requests, there is no benefit in increasing the number of stations per repeater.
    In Appendix~\ref{subsec.MST_kinf}, we discuss this more formally.
\end{itemize}

From this analysis, we conclude that parallel distribution generally fulfills user requests at a higher rate, although in some situations sequential distribution is preferable as it can provide service to a larger number of users.

\subsection{Many users over long distances}\label{subsec.going_far}
Next, we investigate the effect of increasing the number of users and the distances between them.
Here, we assume the same all-photonic forwarding stations from ref. \cite{Niu2022} as in the large-budget use case. This means that the probability of successful packet delivery depends on the distance between users and on the number of repeaters according to~(\ref{eq.p_allphotonic}).
However, as opposed to the large-budget use case, we now consider a fixed but arbitrary number of repeaters per user, $N$, to be placed between the user and the central repeater -- that is, we do not choose $N$ according to any cost function optimization.
Each repeater has a fixed number of forwarding stations~$k$.
The question we address here is: \emph{how far apart can the users be to provide service to them?}

Specifically, we want to analyze the critical distance $L_\mathrm{crit}$ that overloads the system, i.e., the value of $L$ that makes the load from (\ref{eq.load}) equal to 1.
Combining (\ref{eq.mean_service}) and (\ref{eq.load}) and enforcing $L\geq0$, we find that the critical distance is given by
\begin{equation}\label{eq.Lcrit}
\begin{split}
    L_\mathrm{crit} = \max\bigg(0, \frac{ck}{\lambda_0 u(u-1) m} - \frac{c t_\mathrm{fwd}}{2} \Big(2N\\
    + \mathbb{E} \left[ B_\mathrm{n,w,p(L_\mathrm{crit},N),m} \right] \Big) \bigg).
\end{split}
\end{equation}
Recall that $m$ is the number of packets that are sent in each batch ($m=1$ for sequential distribution and $m=k$ for parallel distribution).
Note that (\ref{eq.Lcrit}) is a transcendental equation, since $p$ is a function of $L_\mathrm{crit}$, and it must be solved numerically.

\begin{figure}[t!]
    \centering
    \includegraphics[width=\linewidth]{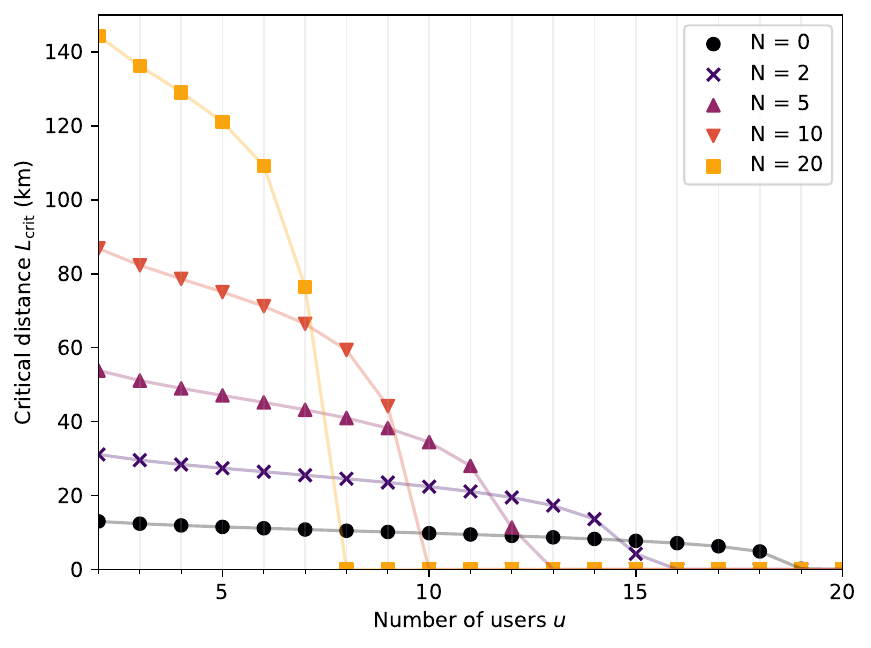}
    \caption{\textbf{Networks with many users cannot cover long distances.}
    Critical distance $L_\mathrm{crit}$ vs number of users $u$, for different numbers of repeaters $N$.
    Parameters used in this figure: sequential distribution, $n=7$, $w\rightarrow\infty$, $k=12$, $\lambda_0 = 10^{-4}$~$\mu$s$^{-1}$, $c=0.2$ km/$\mu$s, $t_\mathrm{fwd} = 100$~$\mu$s.
    }
    \label{fig.going-far}
\end{figure}

The first observation from (\ref{eq.Lcrit}) is that $L_\mathrm{crit}$ decreases (and eventually drops to zero) for increasing $u$, regardless of the number of repeaters $N$.
The second term is always negative, so the critical distance can be upper bounded~as
\begin{equation}\label{eq.Lcrit_bound}
    L_\mathrm{crit} \leq \frac{ck}{\lambda_0 u(u-1) m}.
\end{equation}
The scaling with the number of users is therefore upper bounded by $\sim u^{-2}$.
This unveils a tradeoff between the number of users and the distances between them.
When service is provided to a large number of users, the number of incoming requests will increase and this will put a higher load on the system.
If the distances between users are large, service times will increase, since packets will need to travel further away, which will also increase the load.
Consequently, we must decrease $u$ to increase $L$ and vice versa.
Figure~\ref{fig.going-far} shows an example where we can observe this effect.

We have shown that $L$ and $u$ cannot be scaled up simultaneously when $N$ is fixed, but \emph{what if we add more intermediate repeaters to boost the probability of successful packet delivery? Would this allow us to place many users further apart?}
Intuitively, one may think that this is the case, 
since a larger number of repeaters $N$ provides a larger success probability $p$ (see (\ref{eq.p_allphotonic})).
Increasing $p$ means that we will need to send less (batches of) packets per request, i.e., $\mathbb{E} [B_{n,w,p,m}]$ will be smaller, and consequently the service time (\ref{eq.mean_service}) should decrease and the critical distance (\ref{eq.Lcrit}) should increase.
However, this intuition is not always correct, since each additional repeater introduces an additional delay $t_\mathrm{fwd}$, which directly impacts the service time and the critical distance: both (\ref{eq.mean_service}) and (\ref{eq.Lcrit}) include a linear term in $N$.
These linear terms yield an interesting behavior: the delays introduced by the repeaters accumulate and the service time eventually increases (and the critical distance decreases) when increasing the number of repeaters $N$, despite the benefit from a larger~$p$.
If there are too many repeaters, these forwarding delays may overload the system, preventing requests from being met within a finite amount of time.

When the number of users is small, increasing the number of repeaters is actually beneficial: it allows us to increase the distances between users (i.e., larger $N$ provides larger $L_\mathrm{crit}$), as can be seen in  Figure \ref{fig.going-far}.
When the number of users is large, service is only possible if the number of repeaters is small. That is, when increasing the number of users, $L_\mathrm{crit}$ drops to zero quicker for larger $N$.
As a consequence, for intermediate values of $u$, there is an finite value of $N$ that maximizes the critical distance.
For instance, when there are ten users in the example from Figure~\ref{fig.going-far}, using $N=5$ repeaters allows the users to be further apart than if we used $N=10$ repeaters.
Note that Figure~\ref{fig.going-far} assumes sequential distribution of packets, although the previous discussion remains general and also applies to parallel distribution, as shown in Appendix~\ref{app.going_far}.

From this analysis we draw two main conclusions:
\begin{enumerate}
    \item There exists a tradeoff in the implementation of a QCS protocol: the number of users and the distances between them cannot be scaled up simultaneously.
    This tradeoff happens for any number of repeaters $N$ and any function $p(L,N)$ (see (\ref{eq.Lcrit_bound})), which means that the tradeoff will exist even with unlimited resources and ideal hardware.
    \item Increasing the number of repeaters to boost the probability of successful packet delivery is not necessarily desirable, since we must also consider the forwarding delays.
\end{enumerate}

\section{Outlook}\label{sec.outlook}
This paper lays the groundwork for further exploration and refinement of quantum circuit switching (QCS) protocols in the context of one-way quantum networks.
We have explored two fundamental resource-allocation strategies: in the first one, quantum data packets are distributed sequentially, and, in the second one, they are distributed in batches.
We concluded that sequential distribution can generally provide service to a larger number of users, although parallel distribution generally meets requests at a higher rate.
We also found a tradeoff between the number of users supported by the QCS protocol and the size of the network.

Here, we have considered the mean sojourn time as the main quantity to measure protocol performance. A more detailed characterization of the whole probability distribution of the sojourn time is left as future work.

Future research could also focus on extending the analysis to more complex network topologies beyond the star configuration considered here.
Investigating the impact of connectivity patterns on the performance of QCS protocols would provide valuable insights for real-world quantum network deployment.
Moreover, exploring adaptive QCS strategies that dynamically adjust resource allocation based on traffic conditions (i.e., finding an intermediate strategy between sequential and parallel distribution of packets) could enhance efficiency and scalability.

Different request models must be also investigated in future work. Here, we have assumed that each pair of users submits requests according to a Poisson process. However, some networks may experience different types of requests, e.g., peak demands may happen at specific times of the day.
Moreover, each pair of users could request a different number of packets $n$ to be distributed over a different time window $w$.
Considering a different request model would potentially lead to different conclusions about the scalability of the network (e.g., we would expect a different scaling of the critical number of users with the number of forwarding stations, and a different relationship between critical distance and number of users).

We consider QCS and its integration into one-way quantum networks as a promising avenue towards practical quantum information processing and quantum internet applications.



\section*{Code availability}
The code used to generate all the plots in this paper can be found in the following GitHub repository: \href{https://github.com/AlvaroGI/quantum-circuit-switching}{https://github.com/AlvaroGI/quantum-circuit-switching}.

\section*{Acknowledgment}
We thank B. Davies, D. Towsley, and the Wehner group for discussions and feedback on this project.
We thank B. Davies, L. Prielinger, T. Beauchamp, and G. Vardoyan for critical feedback on this manuscript.

ÁGI acknowledges financial support from the Netherlands Organisation for Scientific Research (NWO/OCW), as part of the Frontiers of Nanoscience program. SW acknowledges support from an NWO VICI grant.

\vspace{20pt}

\bibliographystyle{IEEEtran}
\bibliography{references}

\clearpage
\onecolumn
\appendices
\renewcommand{\theequation}{\thesection.\arabic{equation}}

\section{Analytical calculation of the mean waiting time and the mean service time}\label{app.analytical}
\vspace{10pt}

\begin{figure}[th]
    \centering
    \includegraphics[width=0.5\linewidth]{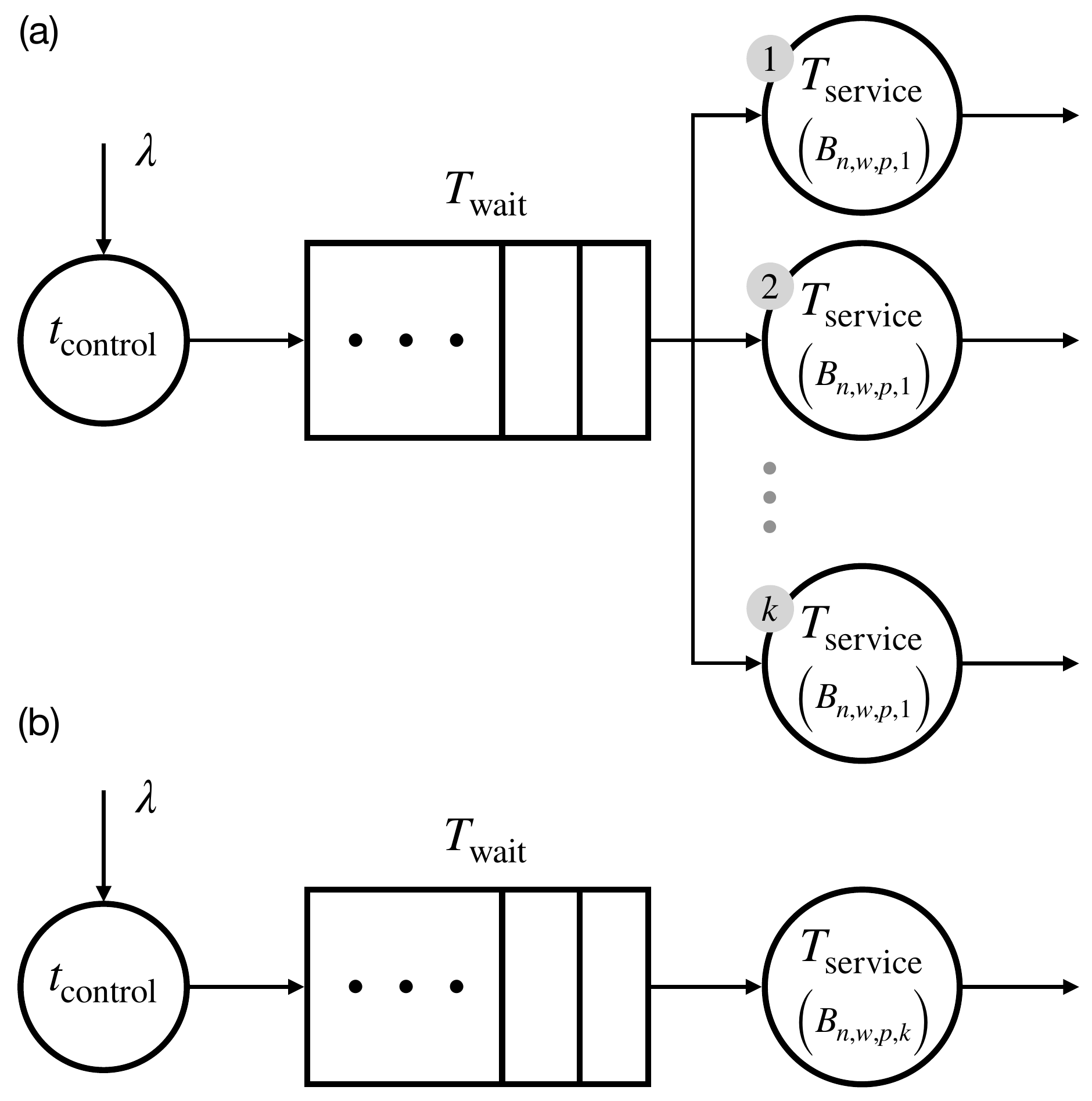}
    \caption{\textbf{QCS in a star network can be modeled as an M/G/$s$ queue.}
    Illustration of the queue model of a QCS scheme with (a) sequential distribution of packets ($s=k$) and (b) parallel distribution of packets ($s=1$), in a star network with $k$ forwarding stations per repeater. New requests are submitted at rate $\lambda$. After some control time $t_\mathrm{control}$, the request is placed in a first-in first-out queue.
    After some waiting time $T_\mathrm{wait}$, the request leaves the queue and begins being processed. Processing the request takes time $T_\mathrm{service}$.
    Sequential distribution (a) can process up to $k$ requests in parallel, while parallel distribution (b) can only process one at a time.
    The service time distribution is given by the distribution of the number of batches of packets that must be sent until the request is met, $B_{n,w,p,m}$ (see Appendix \ref{subsec.servicetime}).}
    \label{fig.queues}
\end{figure}

In this Appendix, we show how to calculate the mean waiting time and the mean service time of a QCS protocol in a star network.
For that, we model the system as an M/G/$s$ queue.
In this queue model, new requests arrive following a Poisson distribution, i.e., request arrivals are Markovian (M).
Incoming requests are placed in a common queue.
Requests in the queue are processed according to a first-in-first-out policy.
Processing a request takes some service time, which follows a general distribution~(G), and a maximum of $s$ requests can be processed simultaneously.
When quantum data packets are distributed sequentially ($m=1$), each of the $k$ forwarding stations is dedicated to meet one request and therefore $s=k$ (see Figure \ref{fig.queues}a).
When packets are distributed in parallel ($m=k$), all forwarding stations are reserved to meet one request at a time and therefore $s=1$ (see Figure \ref{fig.queues}b).

In Sections \ref{subsec.waitingtime} and \ref{subsec.servicetime}, we show how to compute waiting and service times, respectively, in the M/G/$s$ model for QCS.
These results are summarized in Tables \ref{tab.waiting} and \ref{tab.service}.
We conclude this Appendix discussing the limit when the number of forwarding stations goes to infinity (Section \ref{subsec.MST_kinf}).

\renewcommand{\arraystretch}{3}
\begin{table}[h]
    \centering
    \caption{How to compute the mean waiting time in QCS in star networks, depending on the type of packet distribution (sequential or parallel) and the number of forwarding stations per repeater ($k$).
    We assume that the first two moments of the service time distribution are known.
    See Appendix \ref{subsec.waitingtime} for further details.}\label{tab.waiting}
    \vspace{-2mm} 
\begin{tabular}{rlcc}
    & & Sequential ($s=k$) & Parallel ($s=1$)
\\ \cline{3-4}
	$k=1$: & & Closed-form - (\ref{eq.waitexact}) & Closed-form - (\ref{eq.waitexact}) \\
    $k>1$: & & Approximation - (\ref{eq.waiting_approx}) & Closed-form - (\ref{eq.waitexact}) \\
\end{tabular}

\end{table}

\renewcommand{\arraystretch}{3}
\begin{table}[h]
    \centering
    \caption{How to compute the first two moments of $B_{n,w,p,m}$. Once these moments are known, the first two moments of the service time can be trivially computed with (\ref{eq.mean_serviceapp}) and (\ref{eq.mean_service2app}).
    The method used to compute $B_{n,w,p,m}$ depends on the type of packet distribution (sequential or parallel), the window size ($w$), and the probability of successful packet delivery ($p$). See Appendix \ref{subsec.servicetime} for further details.
    }\label{tab.service}
    \vspace{-2mm} 
\begin{tabular}{rlcc}
    & & Sequential ($m=1$) & Parallel ($m=k$)
\\ \cline{3-4}
	$p=1$: & & Closed-form - (\ref{eq.B_p1}) and (\ref{eq.B_p12}) & Closed-form - (\ref{eq.B_p1multi}) and (\ref{eq.B_p12multi}) \\
    $w\rightarrow\infty$: & & Closed-form - (\ref{eq.B_winf}) and (\ref{eq.B_winf2}) & Analytical - (\ref{eq.Pr_B_winf_multi}), (\ref{eq.EB_winf_multi}), and (\ref{eq.EB2_winf_multi}) \\
    $p<1$ and $w<\infty$: & & Analytical but expensive to evaluate - \cite{Davies2023} & Monte Carlo sampling \\
\end{tabular}

\end{table}

\vspace{15pt}
\subsection{Waiting time}\label{subsec.waitingtime}
Calculating the waiting time poses a major problem: no general analytical solution is known for the waiting time of an M/G/$s$ queue.
Only for $s=1$, a closed-form solution is known (see, e.g., Chapter 14 from \cite{VanMieghem2014}):
\begin{equation}\label{eq.waitexact}
    \mathbb{E}\left[T_\mathrm{wait}^{\mathrm{M/G/}1}\right] = \frac{\lambda \mathbb{E}\left[\left(T_\mathrm{service}^{\mathrm{M/G/}1}\right)^2\right]}{2\left(1-\lambda\mathbb{E}\left[T_\mathrm{service}^{\mathrm{M/G/}1}\right]\right)},
\end{equation}
where $\lambda$ is the request arrival rate (in our problem, $\lambda=\lambda_0 u(u-1)/2$, where $\lambda_0$ is the request rate per pair of users and $u$ is the number of users).
The mean waiting time can therefore be computed exactly when $s=1$, provided that the first two moments of the service time are known.

Multiple approximations and bounds for the waiting time distribution have been found for $s>1$ (see, e.g., \cite{Gupta2010}).
One of the most well-known approximations is the early formula from ref.~\cite{Lee1959}:
\begin{equation}\label{eq.waiting_approx}
    \mathbb{E}\left[T_\mathrm{wait}^{\mathrm{M/G/}s}\right] \approx \frac{C_\mathrm{service}^2+1}{2} \mathbb{E}\left[T_\mathrm{wait}^{\mathrm{M/M/}s}\right],
\end{equation}
where $C_\mathrm{service}^2 \equiv \mathrm{Var}\left[ T_\mathrm{service}^{\mathrm{M/G/}s} \right] / \mathbb{E}\left[ T_\mathrm{service}^{\mathrm{M/G/}s} \right]^2$ is the squared coefficient of variation of the service time distribution, and M/M/$s$ is an auxiliary queuing system with a single queue, Markovian arrivals (M), exponentially distributed service times (M) with mean $\mu \equiv \mathbb{E}\left[ T_\mathrm{service}^{\mathrm{M/G/}s} \right]$, and capacity to process $s$ requests simultaneously.
The mean waiting time of the auxiliary queue can be computed as follows (see Chapter 14 from~\cite{VanMieghem2014}):
\begin{equation}
    \mathbb{E}\left[ T_\mathrm{wait}^{\mathrm{M/M/}s} \right] = 
    \frac{\lambda^s}{s! s \mu^{s+1} z^2}
    \left( \frac{\lambda^s}{s! \mu^s z}
    + \sum_{i=0}^{s-1} \frac{\lambda^i}{i! \mu^i}
    \right)^{-1},
\end{equation}
with $z \equiv 1-\lambda/(s\mu)$.
Refs.~\cite{Gupta2010} and \cite{Whitt1993} provide empirical evidence that the approximation (\ref{eq.waiting_approx}) works well when the service time distribution has a small squared coefficient of variation $C_\mathrm{service}^2$. 
We observed empirically that, in the use cases studied in this paper, $C_\mathrm{service}^2$ is indeed small (see Appendix \ref{app.coeff_of_var}) and therefore we expect this approximation to be accurate. Note also that the approximation assumes the first two moments of the service time distribution are known.

To study our QCS protocol, we can employ the closed-form solution (\ref{eq.waitexact}) when ($i$) packets are distributed sequentially and there is a single forwarding station per repeater ($k=1$) and ($ii$) when packets are distributed in parallel.
When packets are distributed sequentially with $k>1$, we can only use approximations, such as (\ref{eq.waiting_approx}), or Monte Carlo sampling.
In practice, the use of both the closed-form solution and the approximation are restricted to situations in which we can efficiently compute the first two moments of the service time. We discuss this in the next section.

\vspace{15pt}
\subsection{Service time}\label{subsec.servicetime}
As introduced in the main text, $B_{n,w,p,m}$ is the number of batches of quantum data packets that must be sent from user to user until $n$ packets are successfully distributed within a time window $w$, assuming each packet has a success probability $p$ and each batch contains $m$ packets.
The service time can be computed as
\begin{equation}\label{eq.serviceapp}
    T_\mathrm{service} = \frac{2L}{c} + t_\mathrm{fwd}(2N+1) + t_\mathrm{fwd} \left(B_{n,w,p,m}-1\right),
\end{equation}
where $L$ is the distance between each user and the central repeater, $c$ is the speed of light in the physical channels, $t_\mathrm{fwd}$ is the forwarding time, and $N$ is the number of repeaters between each user and the central repeater.
The first term corresponds to the time required for the first packet (or batch of packets) to travel from one user to another, without considering forwarding delays. The second term corresponds to the delay introduced by the forwarding stations: each station takes time $t_\mathrm{fwd}$ to forward a packet, and the packet (or batch of packets) is forwarded at $2N+1$ intermediate repeaters.
Therefore, the first two terms account for the total time since the first packet (or batch of packets) is sent from one user until it is received by the other user.
The third term accounts for the time it takes to receive the remaining $B_{n,w,p,m}-1$ (batches of) packets: a packet (or batch of packets) is sent every $t_\mathrm{fwd}$ units of time, to ensure it will arrive at the next repeater when the previous packet is already processed and leaving the repeater.
The mean service time is then given by
\begin{equation}\label{eq.mean_serviceapp}
    \mathbb{E}\left[T_\mathrm{service}\right] = \frac{2L}{c} + t_\mathrm{fwd}(2N+\mathbb{E}\left[B_{n,w,p,m}\right]),
\end{equation}
since $B_{n,w,p,m}$ is the only random variable involved.
As shown in Section \ref{subsec.waitingtime}, the second moment of the service time is also required to compute the expected waiting time $\mathbb{E}[T_\mathrm{wait}]$. The second moment can be computed as
\begin{equation}\label{eq.mean_service2app}
    \mathbb{E}\left[T_\mathrm{service}^2\right] = 
    \left( \frac{4L^2}{c^2} + \frac{8L}{c} t_\mathrm{fwd}N + 4 t_\mathrm{fwd}^2 N^2 \right)
    + \left( \frac{4L}{c} t_\mathrm{fwd} + 4 t_\mathrm{fwd}^2 N \right) \mathbb{E}\left[B_{n,w,p,m}\right]
    + t_\mathrm{fwd}^2 \mathbb{E}\left[B_{n,w,p,m}^2\right].
\end{equation}

From (\ref{eq.mean_serviceapp}) and (\ref{eq.mean_service2app}), we conclude that the service and the waiting times can be efficiently computed if we can efficiently compute the first two moments of $B_{n,w,p,m}$. Next, we discuss how to compute them.

Consider a sequence of i.i.d. random variables $S_i$, which follow a binomial distribution with $m$ attempts and probability of success $p$.
The value of $S_i$ corresponds to the number of successful packets delivered in batch $i$.
Let $B_{n,w,p,m}$ be the number of batches required until completion, i.e.,
\begin{equation}\label{eq.B_definition}
    B_{n,w,p,m} = \inf\left\{x \;:\; \sum_{j=x-w+1}^{x} S_j \geq n\right\}.
\end{equation}
It is possible to calculate the probability distribution of $B_{n,w,p,m}$ analytically when there is no multiplexing ($m=1$) \cite{Davies2023}.
When $m=1$ and $p=1$, the solution is trivial since we deterministically need $n$ batches to successfully deliver $n$ packets (there is only one per batch). In that case, we have
\begin{equation}\label{eq.B_p1}
    \mathbb{E}[B_{n,w,1,1}] = n,
\end{equation}
\begin{equation}\label{eq.B_p12}
    \mathbb{E}\left[ \left(B_{n,w,1,1} \right)^2 \right] =n^2.
\end{equation}

As shown in ref.~\cite{Davies2023}, in the case of $m=1$ and an infinite window ($w\rightarrow\infty$), $B_{n,w,p,1}$ follows a negative binomial distribution, which has the following first two moments:
\begin{equation}\label{eq.B_winf}
    \mathbb{E}[B_{n,\infty,p,1}] = \frac{n}{p},
\end{equation}
\begin{equation}\label{eq.B_winf2}
    \mathbb{E}\left[ \left(B_{n,\infty,p,1} \right)^2 \right] = \frac{n(1-p)+n^2}{p^2}.
\end{equation}

For every other combination of values of $n$, $p$, and $w$, and with $m=1$, there exists a nontrivial analytical solution, as shown in ref.~\cite{Davies2023}.
However, computing the solution involves inverting a matrix whose size scales as $\mathcal{O}(w^{n-1})$.
Hence, in practice we need to employ approximate methods (such as Monte Carlo sampling) for arbitrary values of $n$, $w$, and $p$.

In the multiplexed case ($m>1$), there is no known analytical solution yet for every combination of $n$, $w$, $p$, and $m$, to the best of our knowledge.
Next, we solve the problem for two cases: ($i$) $p=1$ and ($ii$) $w\rightarrow\infty$. In every other case, we employed Monte Carlo sampling to estimate the probability distribution of $B_{n,w,p,m}$ when needed.
First, the case $p=1$ is again trivial. When packets are always successfully delivered, we need $\lceil n/m \rceil$ batches to deliver $n$ packets. Hence,
\begin{equation}\label{eq.B_p1multi}
    \mathbb{E}[B_{n,w,1,m}] = \left\lceil \frac{n}{m} \right\rceil,
\end{equation}
\begin{equation}\label{eq.B_p12multi}
    \mathbb{E}\left[ \left(B_{n,w,1,m} \right)^2 \right] = \left\lceil \frac{n}{m} \right\rceil^2.
\end{equation}

For $w\rightarrow\infty$, we first calculate the survival function of $B_{n,\infty,p,m}$ as follows:
\begin{align}
    \mathrm{Pr}\left[ B_{n,\infty,p,m} > b \right] &\stackrel{a}{=} \mathrm{Pr}\left[ \sum_{i=1}^{b} S_i < n \right]\\
    &\stackrel{b}{=} \sum_{s_b=0}^{m} \mathrm{Pr}\left[ \sum_{i=1}^{b-1} S_i < n-s_b \;|\; S_b=s_b \right] \mathrm{Pr}\left[ S_b=s_b \right]\\
    &\stackrel{c}{=} \sum_{s_b=0}^{m} \mathrm{Pr}\left[ \sum_{i=1}^{b-1} S_i < n-s_b \right] \mathrm{Pr}\left[ S_b=s_b \right]\\
    &\stackrel{d}{=} \sum_{s_1,\dots,s_b=0}^{m} \mathrm{Pr}\left[ \sum_{i=1}^{b} s_i < n \right] \prod_{i=1}^{b} \mathrm{Pr}\left[ S_i=s_i \right]\\
    &\stackrel{e}{=} \sum_{s_1,\dots,s_b=0}^{m} \mathrm{Pr}\left[ \sum_{i=1}^{b} s_i < n \right] \prod_{i=1}^{b} {m \choose s_i} p^{s_i} (1-p)^{m-s_i}\\
    &= \sum_{s_1,\dots,s_b=0}^{m} \mathrm{Pr}\left[ \sum_{i=1}^{b} s_i < n \right] p^{\sum_{i=1}^{b} s_i} (1-p)^{mb - \sum_{i=1}^{b} s_i} \prod_{i=1}^{b} {m \choose s_i}\\
    &= (1-p)^{mb} \sum_{s_1,\dots,s_b=0}^{m} \mathrm{Pr}\left[ \sum_{i=1}^{b} s_i < n \right] \left(\frac{p}{1-p}\right)^{\sum_{i=1}^{b} s_i} \prod_{i=1}^{b} {m \choose s_i}\\
    &\stackrel{f}{=} (1-p)^{mb} \sum_{s_1=0}^{z_1} \sum_{s_2=0}^{z_2} \cdots \sum_{s_b=0}^{z_b} \left(\frac{p}{1-p}\right)^{\sum_{i=1}^{b} s_i} \prod_{i=1}^{b} {m \choose s_i}.\\ \label{eq.Pr_B_winf_multi}
\end{align}
with the following steps:
\begin{enumerate}
    \item[a.] We define $S_i$ as the number of successes in batch $i$. Completing the process in more than $b$ batches ($B_{n,\infty,p,m} > b$) corresponds to obtaining less than $n$ successful attempts in the first $b$ batches ($\sum_{i=1}^{b} S_i < n$).
    \item[b.] We use the law of total probability.
    \item[c.] The number of successes in each batch are independent (i.e., $S_i$ are i.i.d.).
    \item[d.] We apply the previous two steps recursively for every $S_i$.
    \item[e.] $S_i$ follows a binomial distribution with $m$ attempts and probability of success $p$.
    \item[f.] We define $z_i \equiv \min\left(m, n-1-\sum_{j=1}^{i-1} s_j \right)$.
\end{enumerate}

The survival function (\ref{eq.Pr_B_winf_multi}) can then be used to calculate the first two moments of $B_{n,\infty,p,m}$ as follows:
\begin{equation}\label{eq.EB_winf_multi}
    \mathbb{E}[B_{n,\infty,p,m}] = \sum_{b=0}^\infty \mathrm{Pr} \left[B_{n,\infty,p,m}>b\right],
\end{equation}

\begin{equation}\label{eq.EB2_winf_multi}
    \mathbb{E}[B_{n,\infty,p,m}^2] = \sum_{b=1}^\infty b^2 \big( \mathrm{Pr}[B_{n,\infty,p,m}>b-1]
    -\mathrm{Pr} [B_{n,\infty,p,m}>b] \big),
\end{equation}
where we used the fact that $B_{n,\infty,p,m}$ is a discrete and positive random variable.

\vspace{15pt}
\subsection{Limits when $k\rightarrow\infty$}\label{subsec.MST_kinf}
To conclude this Appendix, we show that the mean sojourn time converges to a finite value when the number of forwarding stations, $k$, goes to infinity. For that, we only need to show that the mean waiting time and the mean service time also converge to a constant.

Let us start with the service time. Both $\mathbb{E}[T_\mathrm{service}]$ and $\mathbb{E}[T_\mathrm{service}^2]$ are linear in $\mathbb{E}[B_{n,w,p,m}]$ and $\mathbb{E}[B_{n,w,p,m}^2]$ (see (\ref{eq.mean_serviceapp}) and (\ref{eq.mean_service2app})). Next, we prove that these expected values go to a constant when $k\rightarrow\infty$:
\begin{itemize}
    \item When packets are distributed sequentially, $m=1$, and both $B_{n,w,p,m}$ and $T_\mathrm{service}$ are independent of $k$. Hence, the first two moments are constant in $k$.
    \item When packets are distributed in parallel, $m=k$. When $k\rightarrow\infty$, the number of packets in each batch of the window problem goes to infinity, and therefore the probability of successfully distributing a fixed number of packets $n$ in a single batch goes to 1 (assuming $p>0$, otherwise success is never declared).
    More specifically, the probability that the number of successes in the first batch, $S_1$, is larger than or equal to $n$ goes to 1. This can be proven by showing that this probability is lower bounded by a value that converges to 1:
    \begin{equation*}
    \begin{split}
        \lim_{m\rightarrow\infty} \mathrm{Pr}\left[ S_1 \geq n \right] &=  1 - \lim_{m\rightarrow\infty} \mathrm{Pr}\left[  S_1 < n \right]\\
        &=  1 - \lim_{m\rightarrow\infty} \sum_{x=0}^{n-1} \mathrm{Pr}\left[ S_1 = x \right]\\
        &=  1 - \lim_{m\rightarrow\infty} \sum_{x=0}^{n-1} {m \choose x} p^x (1-p)^{m-x}\\
        &=  1 - \lim_{m\rightarrow\infty}\sum_{x=0}^{n-1} \frac{1}{x!} \left(\frac{p}{1-p}\right)^x \frac{m!}{(m-x)!}(1-p)^{m}\\
        &>  1 - \lim_{m\rightarrow\infty} \sum_{x=0}^{n-1} \frac{1}{x!} \left(\frac{p}{1-p}\right)^x m^x (1-p)^{m}\\
        &=1,
    \end{split}
    \end{equation*}
    where we have used the following: ($i$) $S_i$ follows a binomial distribution with $m$ attempts and probability of success $p$, ($ii$) $m!/(m-x)! < m^x$, and ($iii$) $\lim_{m\rightarrow\infty} m^x q^{m}$ = 0, for $x\geq0$ and $0< q \leq 1$.
    Combining the previous result with (\ref{eq.B_definition}), we obtain $\mathbb{E}[B_{n,w,p,m}]\rightarrow1$ and $\mathbb{E}[B_{n,w,p,m}^2]\rightarrow1$ when $k\rightarrow\infty$.
    Since $\mathbb{E}[T_\mathrm{service}]$ and $\mathbb{E}[T_\mathrm{service}^2]$ are linear in $\mathbb{E}[B_{n,w,p,m}]$ and $\mathbb{E}[B_{n,w,p,m}^2]$ (see (\ref{eq.mean_serviceapp}) and (\ref{eq.mean_service2app})), they also converge to constant values.
\end{itemize}

The waiting time also converges to a constant as $k\rightarrow\infty$:
\begin{itemize}
    \item When packets are distributed sequentially, an infinite number of forwarding stations means that we can simultaneously process as many requests as desired. Intuitively, this means that every incoming request will be immediately processed and the waiting time will be zero. We leave the formal proof as future work.
    \item When packets are distributed in parallel, only one request is processed at a time. The mean waiting time can be computed using (\ref{eq.waitexact}), which only depends on the first two moments of the service time. Since we have shown that these converge to constant values as $k\rightarrow\infty$, it is trivial to show that (\ref{eq.waitexact}) also converges to a constant (assuming $\lambda \mathbb{E} [T_\mathrm{service}^{\mathrm{M/G/}1}] < 1$; otherwise, the system is overloaded and the waiting times go to infinity, as discussed in the main text).
\end{itemize}

\clearpage
\section{Squared coefficient of variation of the service time}\label{app.coeff_of_var}
\vspace{10pt}

As discussed in Appendix \ref{subsec.waitingtime}, there is no known general solution for the expected waiting time of an M/G/$s$ queue, with $s>1$.
Approximations such as (\ref{eq.waiting_approx}) work well when the squared coefficient of variation of the service time, $C_\mathrm{service}^2$, takes small values -- in \cite{Gupta2010}, the authors consider large values of $C_\mathrm{service}^2$ to be in the order of $\gtrsim 10-100$.
In this Appendix, we empirically show that $C_\mathrm{service}^2$ takes small values in systems in which the service time is linear in the number of batches until success of a window problem (this is the case for our QCS protocols).

First, we write the service time as
\begin{equation}\label{eq.servicelinear}
    T_\mathrm{service} = x + y B_{n,w,p,m},
\end{equation}
where $x$ and $y$ are non-negative constants and $B_{n,w,p,m}$ is the number of batches until success, as defined in (\ref{eq.B_definition}). In our work, $x=2L/c + 2 t_\mathrm{fwd} N$, which accounts for the travel time of a quantum data packet from user to user and for the processing delays introduced by the repeaters, and $y=t_\mathrm{fwd}$, which accounts for the delays in between (batches of) packets.
The first and second moments of the service time are
\begin{equation}\label{eq.mean_servicexy}
    \mathbb{E}\left[T_\mathrm{service}\right] = x + y \mathbb{E}\left[B_{n,w,p,m}\right],
\end{equation}
\begin{equation}\label{eq.mean_servicexy2}
    \mathbb{E}\left[ T_\mathrm{service}^2 \right] = x^2 
    + 2xy \mathbb{E}\left[B_{n,w,p,m}\right]
    + y^2 \mathbb{E}\left[B_{n,w,p,m}^2\right].
\end{equation}
For a service time of the form (\ref{eq.servicelinear}), the squared coefficient of variation is upper bounded by the squared coefficient of variation of $B_{n,w,p,m}$:
\begin{equation}\label{eq.C2bound}
\begin{split}
    C_\mathrm{service}^2 &= \frac{\mathrm{Var} \left[T_\mathrm{service}\right]}{\mathbb{E} \left[T_\mathrm{service}\right]^2}\\
    &= \frac{\mathbb{E} \left[T_\mathrm{service}^2\right]}{\mathbb{E} \left[T_\mathrm{service}\right]^2} - 1\\
&= \frac{x^2 
    + 2xy \mathbb{E}\left[B_{n,w,p,m}\right]
    + y^2 \mathbb{E}\left[B_{n,w,p,m}^2\right]}{\left( x + y \mathbb{E}\left[B_{n,w,p,m}\right] \right)^2} - 1\\
&= \frac{x^2 
    + 2xy \mathbb{E}\left[B_{n,w,p,m}\right]
    + y^2 \mathbb{E}\left[ B_{n,w,p,m} \right]^2 \left( 1 + C_{n,w,p,m}^2 \right)}{\left( x + y \mathbb{E}\left[B_{n,w,p,m}\right] \right)^2} - 1\\
&\leq \left( 1 + C_{n,w,p,m}^2 \right) \frac{x^2 
    + 2xy \mathbb{E}\left[B_{n,w,p,m}\right]
    + y^2 \mathbb{E}\left[ B_{n,w,p,m} \right]^2}{\left( x + y \mathbb{E}\left[B_{n,w,p,m}\right] \right)^2} - 1\\
&= C_{n,w,p,m}^2,
\end{split}
\end{equation}
where we have used the fact that $C_{n,w,p,m}^2 \equiv \mathrm{Var}\left[ B_{n,w,p,m} \right] / \mathbb{E}\left[ B_{n,w,p,m} \right]^2 \geq 0$.

\begin{figure}[th]
    \centering
    \includegraphics[width=\linewidth]{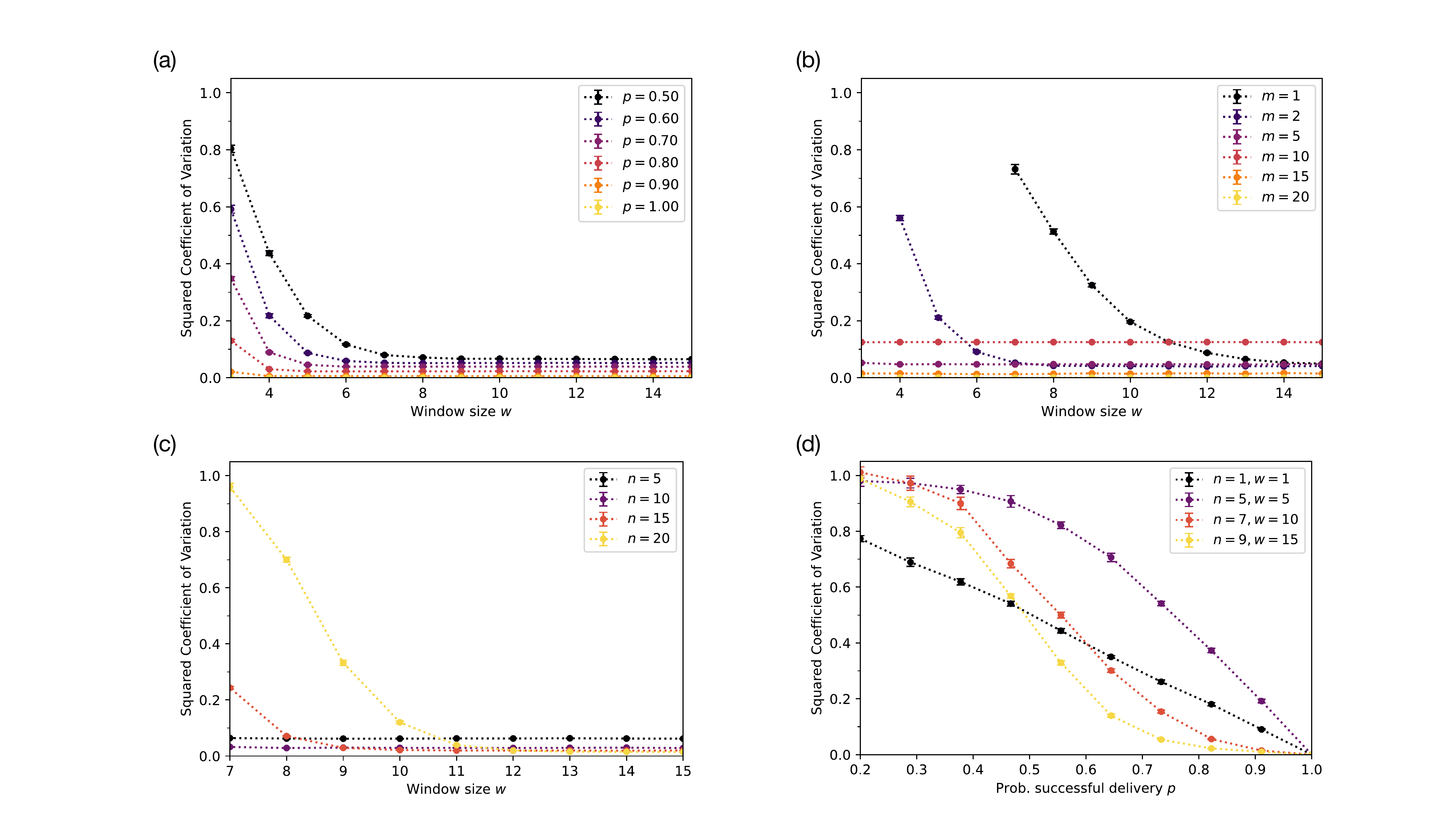}
    \caption{\textbf{The squared coefficient of variation of the window problem is below 1 for the relevant parameter regimes.}
    Squared coefficient of variation of $B_{n,w,p,m}$, $C_{n,w,p,m}^2 \equiv \mathrm{Var}\left[ B_{n,w,p,m} \right] / \mathbb{E}\left[ B_{n,w,p,m} \right]^2$.
    In the parameter regions explored, $C_{n,w,p,m}^2$ decreases with increasing window size $w$ and with increasing probability of success $p$. The behavior with $n$ and $m$ is nontrivial although the values remain below 1 (in the parameter regions explored).
    Each data point is the average over 10 simulations (error bars correspond to the standard error). Each simulation consisted in the estimation of $C_{n,w,p,m}^2$ over $10^3$ samples of $B_{n,w,p,m}$ collected via Monte Carlo sampling.
    Parameters used in each subfigure: (a) $n=7$, $m=3$; (b) $n=7$, $p=0.7$; (c) $p=0.7$, $m=3$; (d) $m=1$.
    }
    \label{fig.sqcoeffvar}
\end{figure}

As a consequence of the bound (\ref{eq.C2bound}), showing that $B_{n,w,p,m}$ has a small coefficient of variation is enough to show that the service time also has a small coefficient of variation.
Figure \ref{fig.sqcoeffvar} shows $C_{n,w,p,m}^2$ for some parameter regimes of $n$, $w$, $p$, and $m$ that are relevant in our work.
This Figure provides empirical evidence that $C_{n,w,p,m}^2$ is below 1 in the parameter regimes explored in this work.
Consequently, according to \cite{Gupta2010}, the mean waiting time should be well approximated by (\ref{eq.waiting_approx}).
Note that there are other interesting features in Figure \ref{fig.sqcoeffvar} (e.g., $C_{n,w,p,m}^2$ is non-monotonic in $n$ and $m$), although we only focus on the order of magnitude of $C_{n,w,p,m}^2$ and therefore a detailed study of the behavior of $C_{n,w,p,m}^2$ is out of the scope of this work.

\clearpage
\section{Parameter values}\label{app.parametervalues}
\vspace{10pt}
In Table \ref{tab.variablesapp}, we provide a summary of all the parameters involved in the model, including the specific values that we use in our examples. Next, we motivate the choice of these numerical values for our examples.

For the number of users $u$, the number of repeaters $N$, and the number of forwarding stations per repeater $k$, we employ values that are reasonable for early quantum networks, on the order of magnitude of $1-10$. Additionally, the specific values we choose provide illustrative examples in which the system exhibits interesting and/or useful behavior (e.g., if we used the same parameter values but the number of users was much larger than 20, the mean sojourn times would diverge to infinity and no service would be possible).
The physical size of the network, $L$, is in the order of magnitude of $1-10$ km. A short distance (1 km) is required for the forwarding stations based on the proposal from ref.~\cite{borregaard2020one}, while all-photonic repeaters are expected to work at internode distances of $1-10$ km \cite{Muralidharan2016, Niu2022}.

The forwarding time of each station $t_\mathrm{fwd}$ is 100 $\mu$s in all our examples. We obtain this value if we consider that: ($i$) approximately ten gates (order of magnitude) are needed to forward a quantum data packet (e.g., the encoding and decoding circuits of the five-qubit perfect code consist of at most 8 gates each \cite{laflamme1996perfect}), and ($ii$) a quantum gate takes around 1-50 $\mu$s (this is the case for qubits realized with color centers, such as nitrogen vacancies -- see, e.g., refs. \cite{Kalb2017, Pompili2021}).
The probability of success $p$ depends on $L$, on $N$, and also on the hardware used to implement the forwarding stations, as discussed in the main text.
We assume the physical channels between users are optical fibers in which the speed of light is approximately $0.2$~km/$\mu$s~\cite{Krutyanskiy2019}.
As discussed in the main text, we consider negligible control time since this would not affect our analysis.

Regarding the requests, increasing the number of requested states, $n$, would increase the expected service time. We tested different values of $n$ and did not find any different behavior from the system.
In our examples, we use $n=7$. We chose this relatively small value of $n$ because a large $n$ would increase the runtime of the Monte Carlo sampling when the analytical solutions cannot be applied.
The request window $w$ must be as large as $n$ (otherwise sequential distribution is not possible and we can only distribute packets in parallel). We consider values of $w$ that are close to $n$ (7, 8, and 10) and also $w\rightarrow\infty$. We do not consider large values of $w$ since the results converge quickly to the infinite window case due to large values of $p$ being used (in most of our examples, $p\geq 0.7$).
Lastly, the value of $\lambda_0 = 10^{-4}$~$\mu$s$^{-1}$ used in our examples was chosen to illustrate interesting behavior, for the other parameter values chosen.
If $\lambda_0$ is small, the system can process requests seamlessly and saturates at a larger number of users (i.e., the critical number of users increases).
If $\lambda_0$ is large, the opposite happens.
Note that, when $\lambda_0 = 10^{-4}$~$\mu$s$^{-1}$, each pair of users submits a request every $10^4$ $\mu$s (on average): during this time, a single forwarding station can forward 100 quantum data packets.

\renewcommand{\arraystretch}{1.6}
\begin{table}[h]
    \centering
    \caption{Parameters of a quantum network running a QCS protocol. In our analysis, we consider a star topology where user nodes are connected via a central repeater, although our definitions and methods remain general. The second column provides the parameter values used in our examples.}\label{tab.variablesapp}
    \vspace{-2mm} 
\begin{tabular}{ccl}
\multicolumn{3}{c}{\textbf{Physical topology}}\\
\hline
	$u$ & $2-20$ & Number of users \\
	$L$ & $1-30$ km & Distance between each user and the central repeater \\
    $N$ & $0-5$ & Number of repeaters between each user and the central repeater\\[5pt]
\multicolumn{3}{c}{\textbf{Hardware}}\\
\hline
	$k$ & $1-15$ & Number of forwarding stations per repeater\\
	$t_\mathrm{fwd}$ & 100 $\mu$s & Forwarding time per repeater and quantum data packet\\
 	$p$ & - & Probability of successful packet delivery from user to user\\
	$c$ & 0.2 km/$\mu$s & Speed of light in the physical channels\\
	$t_\mathrm{control}$ & 0 & Control time\\
\multicolumn{3}{c}{\textbf{Requests}}\\
\hline
	$n$ & 7 & Number of entangled pairs per request \\
	$w$ & $\geq 7$ & Request time window \\
 	$\lambda_0$ & $10^{-4}$ $\mu$s$^{-1}$ &  Request submission rate per pair of users
\end{tabular}

\end{table}

\clearpage
\section{Attenuation in all-photonic quantum repeaters}\label{app.allphotonic}
\vspace{10pt}
In this work, the probability of successfully delivering a quantum data packet, $p$, is assumed to depend on the physical implementation of the forwarding stations, the distance between repeaters ($L_0$), and the number of intermediate repeaters between users.
In some of the use cases discussed in the main text, we consider the all-photonic forwarding stations proposed in ref. \cite{Niu2022}. In this Appendix, we provide more details about how this choice of hardware determines the dependence of $p$ on $L_0$. In particular, we explain how we calculate the effective attenuation coefficient $\alpha_\mathrm{eff}(L_0)$ from~(\ref{eq.p_allphotonic}).

\begin{figure}[th]
    \centering
    \includegraphics[width=0.5\linewidth]{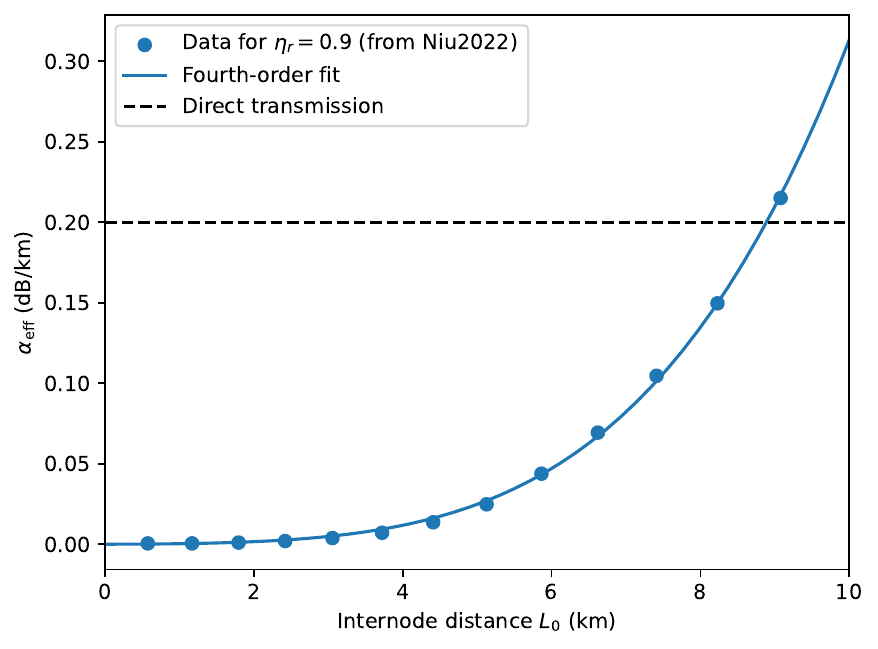}
    \caption{\textbf{Effective attenuation coefficient vs internode distance.}
    We fit $\alpha_\mathrm{eff}$ versus $L_0$ for $\eta_r=0.9$ and the [[48, 6, 8]] code to the data from Fig 4b from ref.~\cite{Niu2022} (blue dots, raw data extracted with WebPlotDigitizer \cite{Rohatgi2022}). Fourth order fitting with Plotly: $\alpha_\mathrm{eff} \approx 10^{-6} (277 L_0^2 + 29 L_0^4)$ dB/km (solid line).
    Direct transmission over optical fiber experiences an attenuation of 0.2 dB/km (dashed line).}
    \label{fig.alpha_eff}
\end{figure}

In our examples, we consider repeaters that employ the [[48, 6, 8]] generalized bicycle code \cite{Panteleev2021}, which was also used as an example in ref.~\cite{Niu2022}.
Moreover, we include photon-source and detector efficiencies, on-chip loss, and coupling losses into a single parameter: the forwarding station efficiency (or transmittance) $\eta_r$. We assume  $\eta_r=0.9$. This value was also used as an example in ref.~\cite{Niu2022}.
In forwarding stations with efficiency $\eta_r=0.9$ that use the $[[48, 6, 8]]$ code, the effective attenuation coefficient can be approximated by
\begin{equation}
    \alpha_\mathrm{eff}(L_0) \approx 10^{-6} (277 L_0^2 + 29 L_0^4) \mathrm{dB/km}.
\end{equation}
We obtained this expression by fitting a fourth order polynomial to the data provided in ref.~\cite{Niu2022} (see Fig. \ref{fig.alpha_eff}).

\clearpage
\section{Critical number of users with probabilistic packet delivery ($p<1$)}\label{app.ucrit_probabilistic}
\begin{figure}[h]
    \centering
    \includegraphics[width=0.9\linewidth]{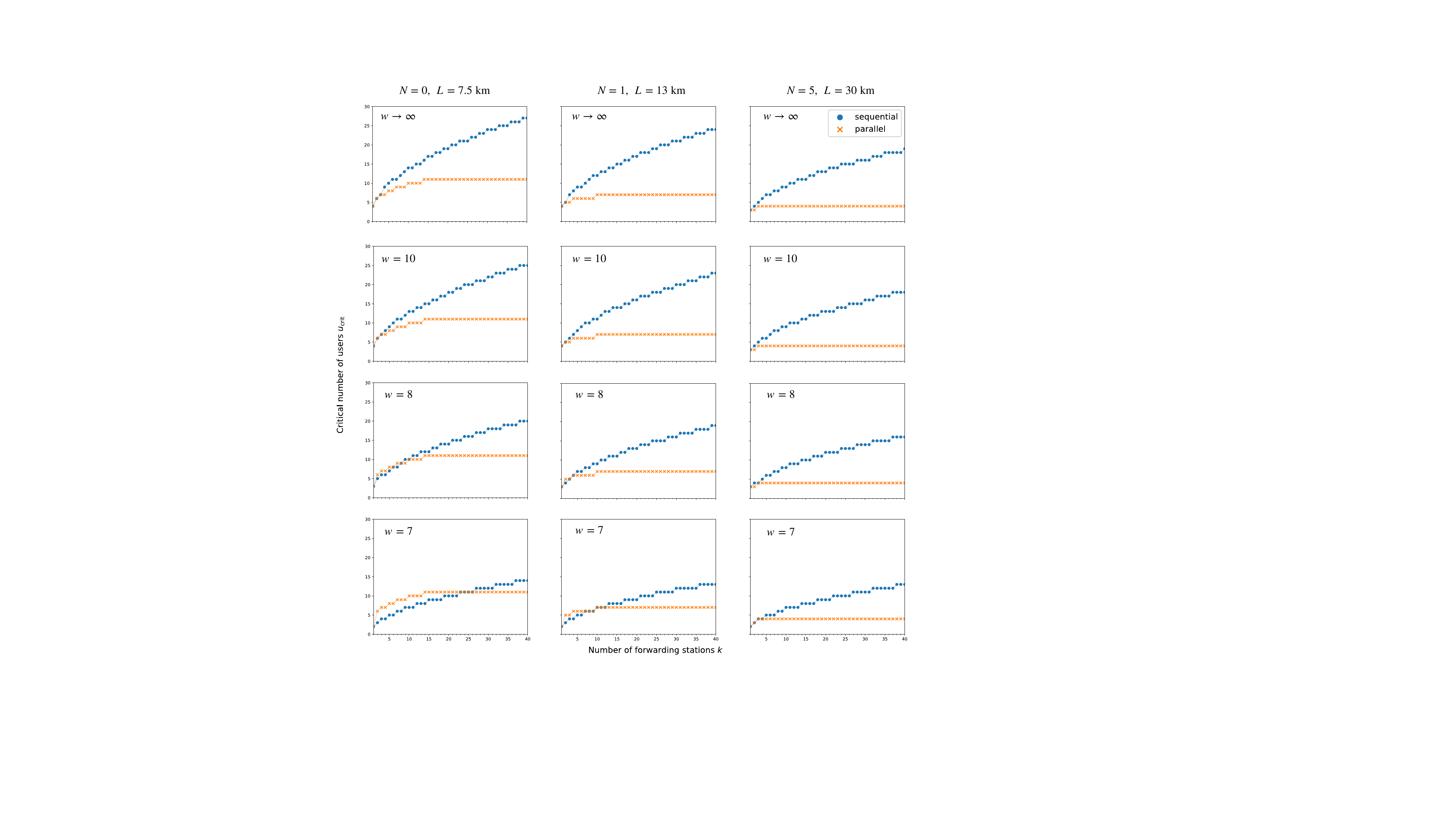}
    \caption{\textbf{Scaling of critical number of users with the number of forwarding stations, for probabilistic quantum data packet delivery.}
	Critical number of users, $u_\mathrm{crit}$, vs number of forwarding stations, $k$, for QCS with sequential (blue dots) and parallel (orange crosses) distribution of packets.
    Star network with $N=0,1,5$ (left to right) repeaters between each user and the central repeater, and distance $L=7.5,13,30$ km (left to right) between each user and the central repeater. These combinations of $N$ and $L$ are cost efficient (they minimize $N/(Lp)$) and yield a probability of successful packet delivery of $p\approx0.7$.
    Nodes request $n=7$ quantum data packets to be successfully delivered within a time window $w=\infty, 10, 8, 7$ (top to bottom).
    When the window size is close to $n$ and only a few forwarding stations per node are available (e.g., bottom left subplot, small $k$), parallel distribution supports more users -- otherwise, sequential distribution supports more users.
    Other parameters used in this figure: $\lambda_0 = 10$~$\mu$s$^{-4}$, $t_\mathrm{fwd} = 100$~$\mu$s.
    Results calculated using (\ref{eq.ucrit}) and (\ref{eq.mean_service}). For $w\rightarrow\infty$, $\mathbb{E}\left[B_{n,w,p,m}\right]$ (required to compute (\ref{eq.mean_service})) was computed using the analytical solution from Appendix \ref{subsec.servicetime} -- in the other cases, the probability distribution of $B_{n,w,p,m}$ was estimated with $10^6$ Monte Carlo samples.
    }
    \label{fig.crit-users-p07}
\end{figure}

\vspace{10pt}
In this Appendix, we provide more examples of the behavior of the critical number of users, $u_\mathrm{crit}$, with increasing number of forwarding stations, $k$, in the large-budget use case from Section \ref{sec.results}.
Figure \ref{fig.crit-users-p07} shows $u_\mathrm{crit}$ vs $k$ for $w=7,8,10,\infty$ and $N=0,1,5$ (corresponding to $L=7.5,13,30$ km). In these cases, the probability of successful packet delivery is $p\approx0.7$. As in the small-budget scenario, $u_\mathrm{crit}$ increases with increasing $k$ for sequential distribution of packets. Conversely, $u_\mathrm{crit}$ reaches a maximum value when packets are distributed in parallel, i.e., we cannot increase $u_\mathrm{crit}$ indefinitely by increasing $k$, for parallel distribution.

As discussed in the main text, parallel distribution supports more users than sequential distribution only when the window size is small (i.e., close to the number of states requested, $n$) and when there are few forwarding stations per repeater (small~$k$).
In Figure \ref{fig.crit-users-p07}, we have $n=7$, and we observe this behavior when $w=7,8$ but not for $w\geq10$.
We also observe that, in the parameter regimes explored, this behavior vanishes when increasing the size of the network: for $L=7.5$ km, parallel distribution can support more users than sequential distribution for $w=7,8$ and small $k$; but for $L=30$ km, parallel distribution cannot support more users for any values of $w$ and $k$.

\clearpage
\section{Mean sojourn time with all-photonic repeaters}\label{app.MST}
\vspace{10pt}
Here, we provide some additional examples of the behavior of the mean sojourn time (MST) in terms of the number of users, $u$, and the number of forwarding stations per repeater, $k$.

\begin{figure}[h]
    \centering
    \includegraphics[width=0.9\linewidth]{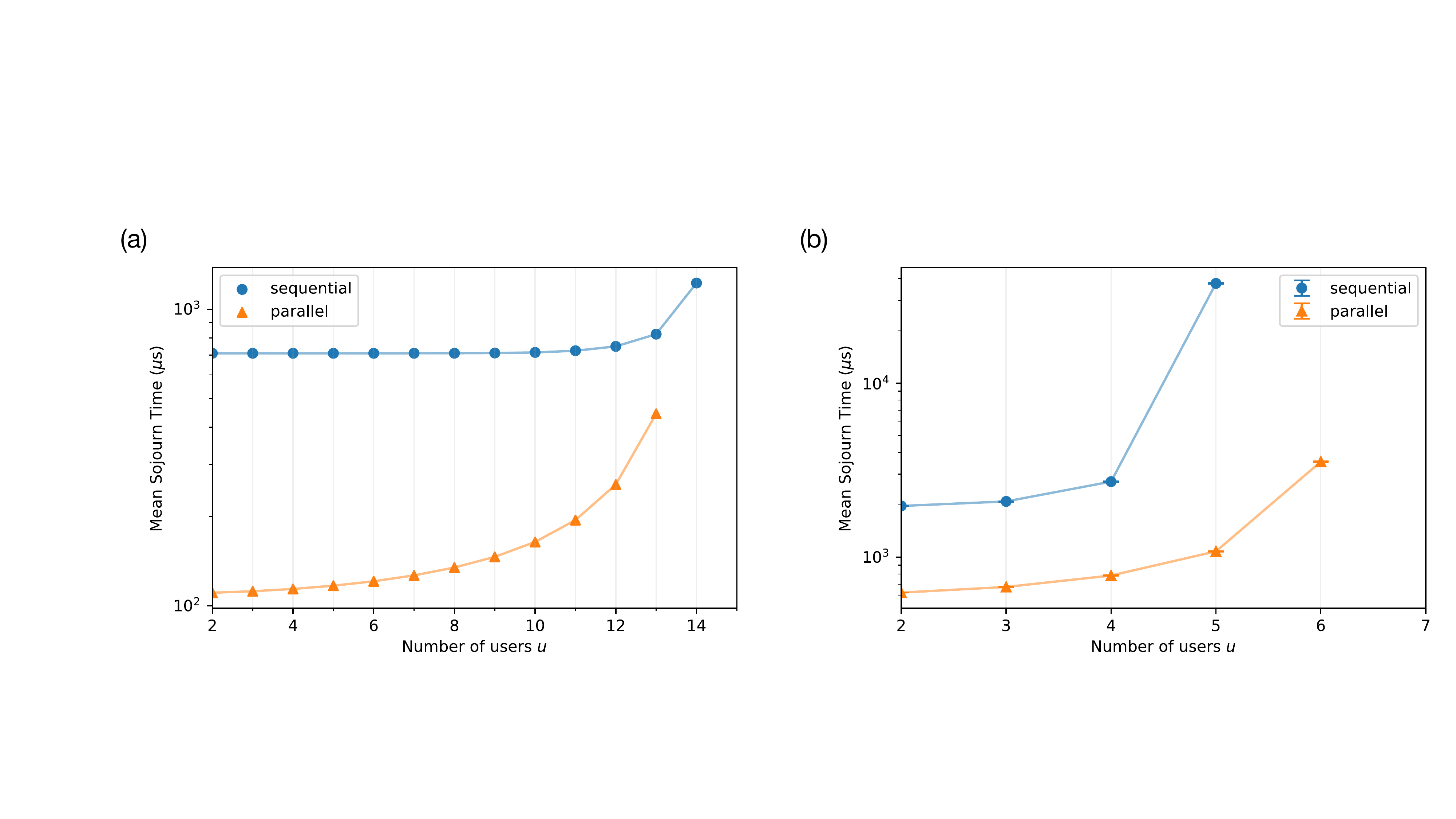}
    \caption{\textbf{Parallel distribution of packets is generally faster, and sometimes it supports more users.}
    MST with sequential (blue) and parallel (orange) packet distribution, for increasing number of users $u$.
    \textbf{(a)} Small-budget use case ($p=1$; $N=0$, $L=1$ km) with $k=7$ and \textbf{(b)} large-budget use case ($p\approx0.7$) with $N=0$, $L=7.5$ km, and $w=8$, and with $k=2$.
    Parameters used in this figure: $n=7$, $\lambda_0 = 10$~$\mu$s$^{-4}$, $t_\mathrm{fwd} = 100$~$\mu$s.
    MST in (a) calculated with (\ref{eq.Tservice_p1}). MST in (b) calculated using a discrete-event simulation and Monte Carlo sampling with $10^5$ samples (the error bars show the standard error).
    }
    \label{fig.MSTvsu-app}
\end{figure}

Figure \ref{fig.MSTvsu-app} shows the MST for sequential and parallel distribution of packets, for increasing number of users and fixed $k$.
In Figure~\ref{fig.MSTvsu-app}a ($p=1$, $N=0$, $L=1$ km, as in Figure~\ref{fig.MST-usecasesAC}a, and $k=7$), sequential distribution can support one more user than parallel distribution (up to 14).
Nevertheless, parallel distribution provides a lower MST, although the advantage decreases as the number of users increases, since we approach the divergence.
In Figure~\ref{fig.MSTvsu-app}b ($p\approx0.7$, $N=0$, $L=7.5$ km, as in Figure~\ref{fig.MST-usecasesAC}b, and $k=2$), parallel distribution can support more users (up to 6, while the MST with sequential distribution diverges after 5 users).
In this case, the advantage in MST provided by parallel distribution actually increases with increasing number of users, since the divergence happens earlier when packets are distributed sequentially.

\begin{figure}[h]
    \centering
    \includegraphics[width=\linewidth]{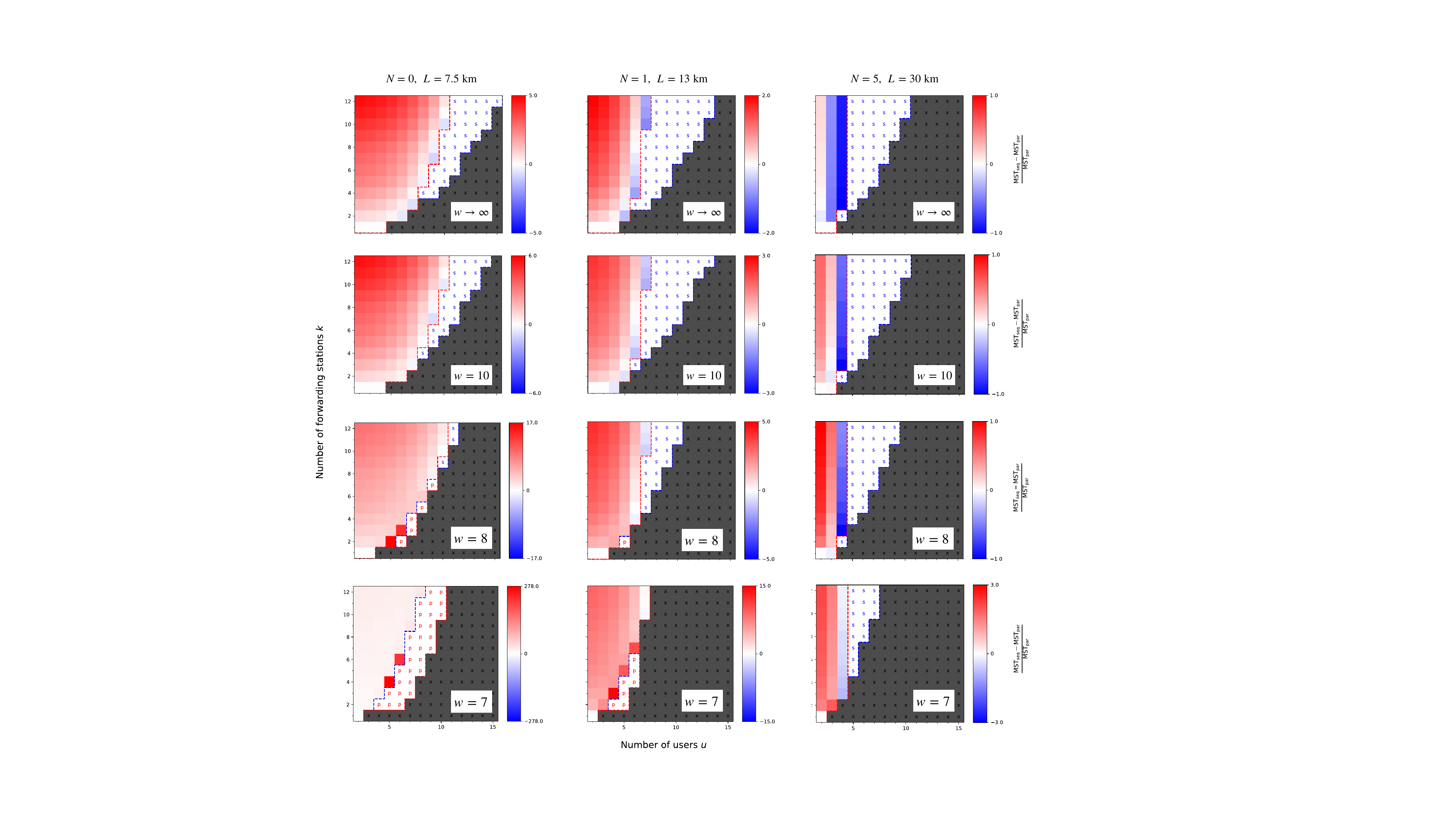}
    \caption{\textbf{Parallel distribution of packets is generally faster.}
    Relative difference in mean sojourn time (MST) between sequential and parallel packet distribution, for different numbers of users $u$ and forwarding stations $k$, in the large-budget use case ($p\approx0.7$) with $N=0,1,5$ and $L=7.5,13,30$ km (left to right), and $w=\infty,10,8,7$ (top to bottom).
    Sequential/parallel distribution provides lower MST in blue/red regions.
    In regions with an `s'/`p', only sequential/parallel distribution can provide service (i.e., yield finite MST).
    In dark regions with an `x', no service is possible.
    Parameters used in this figure: $n=7$, $\lambda_0 = 10$~$\mu$s$^{-4}$, $t_\mathrm{fwd} = 100$~$\mu$s.
    For $w \rightarrow \infty$, MST calculated with (\ref{eq.Tservice_p1}). Otherwise, MST calculated with Monte Carlo sampling with $10^7$ samples (the standard error in the relative difference in MSTs was below 0.5 for every combination of parameters).
    }
    \label{fig.MST-many}
\end{figure}

In Figure~\ref{fig.MST-many}, we provide the difference in MST between sequential and parallel distribution vs the number of users and the number of forwarding stations, for the large-budget use case from the main text.
In this use case, we consider the all-photonic forwarding stations from ref.~\cite{Niu2022}.
The number of intermediate repeaters between each user and the central repeater, $N$, is chosen to minimize the cost $N/(Lp)$. As discussed in the main text, the optimal solution is $p\approx0.7$: for $L=7.5,13,30$ km, we have $N=0,1,5$.
In Figure~\ref{fig.MST-many}, we provide the results for window sizes $w=7,8,10,\infty$.
As discussed in the main text and in Appendix \ref{app.ucrit_probabilistic}, parallel distribution can support more users than sequential distribution only when the resources are scarce, i.e., when $w$ is small compared to $n$ and $k$ is small, and when the network is small. In the figure, this effect happens for $w=7,8$ and $L=7.5,13$ km.

\clearpage
\section{Many users over long distances: sequential vs parallel distribution}\label{app.going_far}
\begin{figure}[h]
    \centering
    \includegraphics[width=0.9\linewidth]{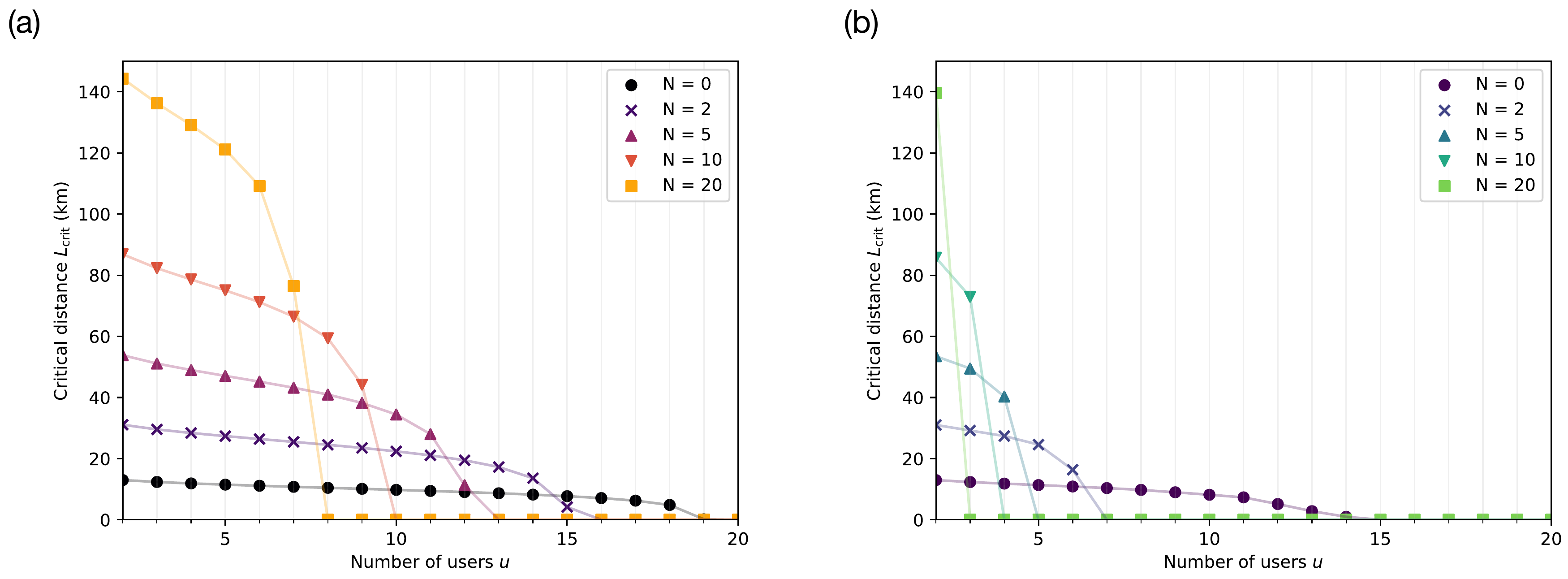}
    \caption{\textbf{Networks with sequential distribution of packets can cover longer distances than with parallel distribution.}
	Critical distance, $L_\mathrm{crit}$, vs number of users, $u$, for different numbers of repeaters $N$ in a star network.
    (a) Sequential distribution of packets; (b) parallel distribution of packets.
    Parameters used in this figure: $n=7$, $w\rightarrow\infty$, $k=12$, $\lambda_0 = 10^{-4}$~$\mu$s$^{-1}$, $c=0.2$ km/$\mu$s, $t_\mathrm{fwd} = 100$~$\mu$s.
    Results calculated using (\ref{eq.Lcrit}), where $\mathbb{E}\left[B_{n,w,p,m}\right]$ (required to solve (\ref{eq.Lcrit})) was computed using the analytical solution from Appendix \ref{subsec.servicetime}.
    }
    \label{fig.lcrit}
\end{figure}

\vspace{10pt}
In this Appendix, we compare sequential vs parallel distribution of packets in terms of the critical distance.
As discussed in Section~\ref{subsec.going_far}, increasing the number of repeaters only increases the critical distance when the number of users is small. Figure~\ref{fig.going-far} shows an example of this phenomenon when using sequential distribution of packets (the same example is shown in Figure~\ref{fig.lcrit}a for convenience).
In Figure~\ref{fig.lcrit}b, we show that the same conclusions are observed when packets are distributed in parallel.
Interestingly, when the number of users is fixed, sequential distribution (Figure~\ref{fig.lcrit}a) allows for larger distances between them (i.e., it provides larger $L_\mathrm{crit}$) than parallel distribution (Figure~\ref{fig.lcrit}b).
We also tested different combinations of parameters ($w=7,8,10$ and $k=6,12,18$) and observed very similar qualitative and quantitative behavior.

\end{document}